\documentclass[letterpaper,twocolumn,10pt]{article}
\usepackage{usenix2019_v3}

\usepackage{amsmath}

\usepackage{booktabs} 
\usepackage{algorithm}
\usepackage{algorithmicx}
\usepackage{algpseudocode}
\usepackage{amsfonts}
\usepackage{amssymb}
\usepackage{amsthm}
\usepackage{pifont}
\usepackage{graphicx}

\usepackage{layouts}

\usepackage{makecell}
\usepackage{tcolorbox}
\usepackage{enumitem}

\usepackage{siunitx}
\usepackage{placeins}
\usepackage{outlines}
\usepackage{svg}

\usepackage{tikz}
\usepackage{pifont}

\usepackage{xspace}

\usepackage[skip=4pt]{caption}
\usepackage{subcaption}
\newtheorem{theorem}{Theorem}[section]
\newtheorem{lemma}[theorem]{Lemma}
\newtheorem{corollary}[theorem]{Corollary}

\newcommand{\prot}{Aspen}

\newcommand{\repair}[0]{\textsc{Repair}}
\newcommand{\alignp}[0]{\textsc{Align}}
\newcommand{\syncp}[0]{\textsc{Sync}}

\newcommand{\kp}[0]{k_{\mathrm{P}}}
\newcommand{\kc}[0]{k_{\mathrm{C}}}
\newcommand{\rd}[0]{r_{\mathrm{D}}}

\newcommand{\signed}[2]{\langle #1 \rangle_{{#2}} }

\newcommand{\p}[0]{p}

\newcommand{\cmark}{\ding{51}} % check mark
\newcommand{\xmark}{\ding{55}} % cross mark

\newcommand*\circled[1]{\tikz[baseline=(char.base)]{
            \node[shape=circle,draw,inner sep=0pt] (char) {#1};}}

\newtcolorbox{takeawaybox}{%
  colback=black!3, colframe=black!45, boxrule=0.4pt, sharp corners,
  left=4pt, right=4pt, top=3pt, bottom=3pt, before skip=4pt, after skip=6pt}

\newcommand{\allnotes}[1]{}
\renewcommand{\allnotes}[1]{#1} % Comment this line out to turn off notes.

\newenvironment{parafont}{\fontfamily{ptm}\selectfont}{}
\newcommand{\Para}[1]
{\noindent\begin{parafont}\textbf{\textit{#1}}\end{parafont}}

\def\fullappendix{}

\begin{document}

%don't want date printed
\date{}

\title{\Large \bf \prot{}: Making Leaderless BFT Fast Paths Practical with Synchronized Clocks}

% Anonymous for double-blind review; use the commented block for camera-ready.
% \author{Paper \#689 \quad (Traditional Research Track)}
\author{
{\rm Daniel Qian}\\
New York University
\and
{\rm Xiyu Hao}\\
New York University
\and
{\rm Jinkun Geng}\\
Stony Brook University
\and
{\rm Yuncheng Yao}\\
New York University Shanghai
\and
{\rm Aurojit Panda}\\
New York University
\and
{\rm Jinyang Li}\\
New York University
\and
{\rm Anirudh Sivaraman}\\
New York University
} % end author

\maketitle

% !TeX root = ../paper.tex

%-------------------------------------------------------------------------------
\begin{abstract}
%-------------------------------------------------------------------------------
No Byzantine Fault Tolerant (BFT) protocol can commit a request in less than the single
round trip it takes for clients to reach the replicas and hear back. Some protocols approach this
bound with a leaderless, speculative fast path, where clients broadcast requests directly
to replicas and commit in two message delays ($2\Delta$). However, such a fast path is
extremely fragile: when clients submit requests concurrently, replicas receive them in
different orders and end up in different states, forcing the protocol into costly recovery.

This paper presents Aspen, a leaderless speculative BFT protocol that can maintain its fast path
even while clients submit requests concurrently, at a near-optimal latency of $2\Delta + \epsilon$.
The $\epsilon$ term is a short waiting delay that Aspen adds to give requests a best-effort,
tentative initial order using loosely synchronized clocks and network delay estimates. Since real networks are not perfectly predictable,
Aspen also tolerates replicas diverging: it adds extra replicas
($n = 3f + 2p + 1$) so that the fast path survives up to $p$ diverged replicas, a lightweight alignment
subprotocol that returns those replicas to the fast path in the background, and a repair
subprotocol that restores agreement when too many diverge.

In experiments with geo-distributed replicas, Aspen reduces
the latency of requests by $1.1\times$--$5.5\times$ compared to
state-of-the-art BFT protocols, while maintaining throughput matching or exceeding throughput-optimized
designs.

\end{abstract}

% !TeX root = ../paper.tex

\section{Introduction}

Byzantine Fault Tolerant (BFT) State Machine Replication (SMR) protocols
\cite{schneider_smr_1990} are a building block for resilient
software systems that can tolerate arbitrary node behavior. They underpin a wide
variety of systems, including blockchains and distributed ledgers
\cite{hyperledgerfabric, russinovich2019ccf, mazieres2015stellar, ethereum, sui},
decentralized computing platforms \cite{icc}, and fault-tolerant 
coordination services \cite{upright, depspace}.

However, the latency overheads of BFT mean that BFT protocols are
not typically deployed on the critical path of user-facing applications. Instead,
practitioners often rely on weaker crash fault tolerance protocols~\cite{hyperledgerfabric}
or secure enclaves~\cite{russinovich2019ccf}. In this work, we focus on addressing the 
latency overhead of BFT and investigate: \emph{how close can a practical BFT
protocol get to responding to clients in a single round trip?}

Our starting point is a family of speculative leaderless protocols, historically referred to as
``Byzantine Quorum'' protocols \cite{QU,700bftaliph,HQosdi}. These protocols achieve a single
round trip latency in their fast path by having a client directly contact all replicas
instead of going through a designated leader. Furthermore, their messaging and cryptographic 
overheads are also lightweight, giving them the potential for high throughput. In practice,
however, this potential is undermined by the brittleness of their fast paths. 
Because clients submit requests directly to replicas without coordination,
replicas must independently observe the same request order. But when two (or more) clients submit
requests concurrently, replicas are likely to see different orders, or \textit{diverge}, 
forcing these protocols to roll back state and commit requests through expensive
recovery paths.

In this work, we present \prot{}, \emph{a speculative leaderless BFT SMR protocol whose fast path
can gracefully handle concurrent requests}.
The main idea behind \prot{} is to use synchronized clocks to let clients set the initial order of
requests without a leader. This design exploits the relative stability 
of cloud networks, along with the increasing availability of clock synchronization 
in both software~\cite{huygens,aws-clock-sync,chrony} and hardware~\cite{spanner,
osdi20-sundial,nsdi22-graham,sigcomm25-firefly,whiterabbit,vldb25-k2}.
At a high level, a client broadcasts requests to replicas annotated with an 
\emph{estimated time of arrival} (ETA),  the time by which a client estimates its 
request will arrive at \textbf{all} replicas. These ETAs are computed from periodic 
network probes that estimate client-to-replica one-way delays (OWDs). Replicas process 
each request only when their local (synchronized) clocks reach the ETA, rather than upon request arrival. 
If requests arrive before their ETAs at  all replicas, then replicas will process requests 
in a common order (i.e., by ETA) and can then commit requests in the fast path. Effectively,
waiting for ETAs adds a short waiting delay to the fast path, but in return, it allows
\prot{} to handle concurrent, correctly behaving clients while clocks are synchronized
and the network is stable.

However, real networks are not perfectly predictable. Queuing delays and path
changes can cause requests to arrive late relative to their ETAs. When a replica receives
a late request, it may have already processed other requests with lower ETAs.
%As a result, if a request arrives late on one replica, its state will diverge from the replicas that received
%requests on time. 
When requests are late, the order in which requests are processed
effectively reverts to being dependent on the timing of message arrivals.

Thus, to be practical in real networks, \prot{} is designed to operate under occasional divergence
of replica state. In particular, \prot{} uses a tunable number of $2p$ extra replicas (for a total of
$n = 3f + 2p + 1$ replicas), which allows up to $p$ replicas to diverge while the rest of the
replicas can continue to make progress in the fast path. These diverged replicas can
rejoin the fast path in the background through a lightweight \textit{alignment} subprotocol that
also relies on ETAs, allowing it to run with little coordination and without interrupting the other
replicas. Only when more than $p$ replicas diverge simultaneously does
\prot{} invoke a heavyweight \textit{repair} subprotocol that interrupts the fast path. 
The repair subprotocol is designed so that \prot{} can return immediately to the fast path 
if the disruption is temporary, while still making progress if the disruption is long lasting.
Together, these mechanisms allow \prot{} to take the fast path most of the time, even when the
network behaves unpredictably. 

In our evaluations, \prot{} can be tuned to remain in the fast path $97\%$--$100\%$ of the time. This
allows \prot{} to improve end-to-end latency by 1.1×–1.3×
compared to PBFT~\cite{pbft} and low-latency protocols such as Zyzzyva~\cite{zyzzyva} and Flutter~\cite{flutterblink}, while sustaining 1.6×–4.0× their throughput. Compared to Autobahn-RSM, a throughput-optimized protocol based on Autobahn~\cite{autobahn} with stronger client verification, \prot{} improves latency by $5.5\times$ while maintaining similar throughput.

In this paper, we make the following contributions:

\begin{itemize}[nosep]
    \item We introduce \prot{}, a speculative, leaderless BFT SMR protocol that handles concurrent clients by ordering requests by ETAs.
    \item We demonstrate how \prot{} gracefully handles intermittent
    replica divergence in the fast path.

    \item We design a repair protocol that preserves safety and liveness in the worst case,
    while also providing reasonable performance when heavily utilized.
\end{itemize}

% !TeX root = ../paper.tex

\section{Background}\label{sec:background}

We compare Byzantine replication protocols by their
\emph{end-to-end latency}: the time from when a client submits a request
until it can safely use the result, expressed in terms of one-way 
message delays ($\Delta$). The trivial lower bound is $2\Delta$ (i.e., a single
round trip): the client must send its request to replicas and wait for the 
execution result. This bound counts message delays alone, and assumes replicas respond
to requests instantaneously, ignoring processing time. However, some
protocols deliberately delay processing requests based on a timer, including
\prot{}, which we denote as adding an $\epsilon$ term to the latency~\cite{flutterblink}.

Many low-latency BFT protocols use \textit{fast path quorums}
\cite{fab, zyzzyva, banyan}.
Specifically, a protocol with $n=3f+2p+1$ replicas, where $p\geq 0$, 
can use a fast path quorum of $(n-p)$ replicas to commit requests faster.
Given a proposal from a correct replica (typically the leader), replicas 
can agree on and decide the proposed value within $2\Delta$ latency
\cite{goodcaselatency, 5f-1}.
However, these bounds cover only the replica-to-replica portion of the
latency. Achieving a $2\Delta$ \emph{end-to-end} latency additionally
requires addressing how client requests are proposed, and when execution
results are returned to clients.

The first requirement for a $2\Delta$ end-to-end latency is \textit{leaderless proposals}.
In traditional leader-based Byzantine replication protocols such as PBFT~\cite{pbft},
a client submits its request to a designated leader, which then proposes the request
to the replicas. The leader simplifies ordering by providing correct replicas with
a common sequence of proposals, but adds an extra hop to the end-to-end client path.
Leaderless protocols~\cite{flutterblink,HQosdi,QU,700bftaliph} remove this hop by
allowing clients to propose requests directly to replicas, but must establish a
common request order without relying on a leader.

The second requirement is \textit{speculative execution}. In typical SMR, replicas
first commit a request, then execute it and send the result to the client.
In speculative protocols such as Zyzzyva~\cite{zyzzyva}, replicas execute
requests before they commit and return speculative results to the client,
which checks them for consistency before delivering. Speculation removes
a message delay by overlapping execution with agreement, but may require
rolling back state if replica logs diverge.

Omitting either property results in a minimum $3\Delta$ end-to-end path.
For example, Zyzzyva is speculative but leader-based: the client sends its
request to the leader, the leader forwards it to replicas, and replicas
speculatively execute and reply, giving a $3\Delta$ fast path. By contrast,
Flutter is leaderless but not speculative: clients send requests directly
to replicas, but replicas use deadlines and communication between
replicas to coordinate on a committed order before execution, yielding a
$3\Delta+\epsilon$ end-to-end path.

Combining leaderless proposals with speculative execution can reduce the
end-to-end path to $2\Delta$~\cite{QU,700bftaliph,HQosdi}, but leaves no
time for replicas to wait or coordinate on an order before replying. To
stay within the $2\Delta$ budget, replicas must therefore speculatively
execute requests as they arrive. Concurrent requests from different clients
may then arrive in different orders at different replicas, causing their
speculative states to diverge and requiring an expensive recovery procedure,
during which no new requests commit on the fast path. \prot{} instead uses
clock-based ETAs to provide a tentative order for speculative execution,
trading a small waiting delay $\epsilon$ for a best-effort common request
order.
% !TeX root = ../paper.tex
%-------------------------------------------------------------------------------
\section{\prot{} Overview}\label{sec:overview}
%-------------------------------------------------------------------------------
\Para{System Model.}
The \prot{} protocol consists of $n = 3f + 2p + 1$ replicas and a set of known clients. We assume there exist at most $f$ faulty (i.e., Byzantine) replicas, but allow for an arbitrary number of faulty clients. Faulty clients and replicas can behave arbitrarily and coordinate, but cannot break standard cryptographic assumptions. Clients and replicas communicate through reliable, authenticated channels. For authentication, we assume the system is set up with secure digital signatures: $\signed{m}{r}$ indicates a message $m$ was signed by process $r$. \prot{} further assumes a partially synchronous network model~\cite{partialsynchrony}, where the network alternates between periods of timely delivery of messages and asynchrony. \prot{} always guarantees safety and provides liveness during periods of synchrony. 

In addition to these standard BFT assumptions, \prot{} requires that the underlying state machine support speculative execution and rollback.
Fast-path execution in \prot{} also requires the client to be correct and to accurately 
estimate the message delay to replicas. 
Clock synchronization is necessary to meet these requirements, and we assume that
clock synchronization is provided as part of the network fabric. 
Note that while Byzantine clients or replicas and poorly synchronized clocks can degrade the performance of \prot{}, they cannot affect safety or liveness. This caveat is similar to prior approaches~\cite{geng2023nezha,sosp25-tiga,epaxostoq} that use clock synchronization for performance, but not correctness~\cite{liskov_clock_sync}. We defer discussion of how to cope with the performance impact of Byzantine processes and clocks to future work.

%%%%%%%

\Para{Clients.}
As with any leaderless protocol, clients in \prot{} have more responsibility than typical SMR clients. Clients must probe the network to estimate their message delays to replicas, set an ETA on each request based on these measurements, then multicast the request to all replicas. Then, due to the speculative nature of \prot{}, they must also collect and check the consistency of replies before delivering the state machine execution result. Specifically, clients can either receive $(n - p)$ consistent speculative replies from the fast path, or $(f + 1)$ committed replies from \repair{}, either of which indicates the request is committed and safe to deliver.

\begin{figure}[!tbp]
    \centering
    \usetikzlibrary{arrows.meta}%
%
% Dark enough to carry 7 pt small caps. Labels are black; grey is kept for
% the \S refs alone, which a reader skips on first pass.
\providecolor{alignink}{HTML}{2B5573}%
\providecolor{repairink}{HTML}{7F4424}%
\providecolor{edgeink}{HTML}{5C5C5C}%
%
% \S refs, so a reader can go straight from a box to the subsection that
% specifies it. Defined out here: a definition inside the picture is not in
% scope when TikZ typesets the nodes.
%
% `hidelinks` first, and the whole thing in a group: acmart turns on
% `colorlinks` with `linkcolor=ACMPurple`, which paints the number but not
% the brackets around it, so the ref comes out purple inside grey parens.
% The link still works, it just takes the colour set here. Guarded because
% the file has to typeset without hyperref too.
%
% preview_tikz.py cannot show this either way: its stub labels carry no
% hyperref anchor, so no link is made and the ref is grey there whatever
% this does. Check the colour in the built paper.
\providecommand{\statesecref}[1]{{%
  \ifdefined\hypersetup\hypersetup{hidelinks}\fi
  \color{edgeink}(\S\ref{#1})}}%
\begin{tikzpicture}[
    font=\fontsize{7}{8.4}\selectfont,
    state/.style={
      draw, rounded corners=0.04in, align=center, inner sep=6pt,
      minimum width=0.78in, minimum height=0.5514in, line width=0.8pt},
    flow/.style={-{Stealth[length=4.4pt, width=2.8pt]}, line width=0.9pt},
    cond/.style={black, align=center, inner sep=2.2pt},
  ]

  % The two things this state does are specified in two different
  % subsections, so each takes its own \S ref and the box takes none: a
  % single ref on the box would have pointed at one of them and quietly
  % dropped the other.
  \node[state] (spec) at (0in, 0in)
    {\textbf{Fast path}\\[2.5pt]
     release from \textit{etaQ} \statesecref{sec:protocol:etas}\\
     periodically \textsc{Sync} \statesecref{sec:checkpointing}};
  \node[state, draw=alignink, text=alignink] (align) at (0in, -1.3125in)
    {\textsc{Align}\\[1.5pt] \statesecref{sec:protocol:align}};
  \node[state, draw=repairink, text=repairink] (repair) at (1.843in, 0in)
    {\textsc{Repair}\\[1.5pt] \statesecref{sec:protocol:repair}};

  % The vertical leg: the Align round trip. Not `sloped`: on a downward path
  % TikZ turns the node to keep the type upright but leaves the line order
  % alone, so a two-line label comes out reversed. Rotating explicitly and
  % anchoring by an edge fixes the reading direction and puts each label
  % outside the pair -- `anchor=south` throws the type left of its arrow,
  % `anchor=north` right of it -- so the two never share the channel between
  % them.
  %
  % On end, a line's length is read down the leg rather than across it, so
  % the leg has to be as long as the longest line and the breaks are chosen
  % to shorten that line, not to balance the block. `Receive conflicting`
  % takes three lines for that reason.
  \draw[flow, alignink]
    ([xshift=-0.085in]spec.south) -- ([xshift=-0.085in]align.north)
    node[cond, midway, rotate=90, anchor=south]
      {Receive\\ conflicting\\ \texttt{CHECKPOINT}};
  \draw[flow]
    ([xshift=0.085in]align.north) -- ([xshift=0.085in]spec.south)
    node[cond, midway, rotate=90, anchor=north]
      {Receive\\ \texttt{STATE-REPLY}};

  % The horizontal leg: the Repair round trip.
  \draw[flow, repairink]
    ([yshift=0.085in]spec.east) -- ([yshift=0.085in]repair.west)
    node[cond, midway, above]
      {\texttt{CONFLICT} /\\ \texttt{TIMEOUT}};
  \draw[flow]
    ([yshift=-0.085in]repair.west) -- ([yshift=-0.085in]spec.east)
    node[cond, midway, below]
      {Finish \textsc{Repair},\\ move to round $i+1$};

  % The hypotenuse: Align to Repair, on the same proofs but a later slot --
  % which is why it reads like the leg above it. Edge to edge rather than
  % corner to corner: corner to corner runs into the return leg's label on
  % its way, and lands its head in the same place the label ends.
  \draw[flow, repairink] (align.east) -- ([xshift=0.10in]repair.south)
    node[cond, midway, sloped, below]
      {\texttt{CONFLICT} / \texttt{TIMEOUT}\\ (later slot)};

\end{tikzpicture}%
    \caption{Replica states.}
    \label{fig:states}
\end{figure}

\Para{Replicas.}
The replicas in \prot{} revolve around their \emph{log}. On each replica, some prefix of the log is committed and guaranteed to be consistent, while the rest of the log is speculative. The speculative portion of the replica logs can diverge if ETAs fail to give replicas a consistent request order, which can happen due to unpredictable networks or faulty clients. We say that two replica logs do not \textit{conflict} if they are equal or if one log is a prefix of the other. We define a \emph{fast path quorum} as a set of $(n - p)$ replicas such that any two replica logs within the quorum do not mutually conflict. A correct replica is \emph{diverged} if it does not belong to such a quorum, which may or may not exist.

Figure~\ref{fig:states} shows the three states a replica moves through: the \emph{fast path}, where replicas speculatively execute requests and clients commit them; \alignp{}, where a diverged replica attempts to rejoin the fast path quorum; and \repair{}, where replicas agree on and commit a new log when no fast path quorum exists. \prot{} proceeds in a series of \emph{rounds} of the fast path separated by invocations of \repair{}. Effectively, this partitions each replica log into sections committed by each round of \repair{}, with the last section including a speculative portion. We describe each state in turn.

\Para{Fast Path.}
Replicas place incoming requests in a priority queue ordered by ETA, and release the request at the head of this queue once their local clock passes its ETA. To process a request, a replica appends it to its log, speculatively executes the state machine operation, and replies to the client with the execution result and a hash of its log. If requests arrive at all replicas before their ETAs, correct replicas process them in the same order and the client commits in the fast path. This fast path is simple and lightweight: clients commit in two message delays plus a small waiting time, and replicas perform no coordination on individual requests.

Replicas also periodically \syncp{} their state by broadcasting the hash of their logs. If a replica gathers a fast path quorum of $(n - p)$ matching hashes, it forms a checkpoint that commits the log up to that point, and notifies other replicas. The checkpoint is also optimistic, only requiring replicas to coordinate on the log digests at specific indices rather than each log entry, effectively batching replica coordination across many requests. However, \syncp{} may not succeed: a replica may find that other replicas created a checkpoint that conflicts with its own log, triggering \alignp{}. Replicas may also time out or conclude no fast path quorum can exist to form a checkpoint, triggering \repair{}. Syncing frequently lets replicas detect these cases quickly, and limits the size of the uncommitted log that \repair{} must agree on.

\Para{Alignment.}
Up to $p$ correct replicas may diverge even if a fast path quorum exists. Such a replica realizes it has diverged when it observes \texttt{CHECKPOINT} messages that conflict with its log, and runs \alignp{} to rejoin the fast path quorum in the background. \alignp{} starts with a state transfer that resets the replica's state to the checkpoint. However, \alignp{} then runs into a unique problem: to rejoin the fast path, the replica must infer the order the quorum used for requests after the checkpoint. It turns out there is a simple solution that reuses the ETAs assigned by clients: the replica replays uncommitted requests by ETA, which works since the quorum is most likely also committing requests in ETA order. \alignp{} is best effort; a replica can fail to rejoin the fast path quorum, though safety cannot be compromised. If there are no faulty clients and the network returns to being stable after the initial divergence, \alignp{} is likely to succeed.

\Para{Repair.}
If more than $p$ replicas diverge or are faulty, creating a checkpoint is impossible, and so replicas will perform the \repair{} subprotocol. \repair{} is a more expensive recovery procedure based on PBFT that interrupts the fast path. Replicas enter \repair{} from either the fast path or \alignp{} by timing out (\texttt{TIMEOUT}) or by collecting enough evidence during \syncp{} to show that a fast path quorum is impossible (\texttt{CONFLICT}). Replicas then share their speculative logs and agree on a new log, preserving any requests previously committed in the fast path while also committing a batch of new requests. Replicas will also send the clients replies indicating which requests are committed. Once \repair{} succeeds, replicas move to the next round and return to the fast path. However, \repair{} is expensive, requiring the entire uncommitted history to be shared, and client requests that arrive immediately before and during \repair{} experience elevated latency waiting for it to finish.

The safety and liveness of \prot{} ultimately rely on \repair{}. Safety is ensured since \repair{} guarantees replicas agree on logs consistent with the fast path and previous rounds. Liveness under partial synchrony is provided because \repair{} can commit requests even if the fast path is unable to. Appendix~\ref{sec:appendix:proof} gives full proofs of safety and liveness.
% !TeX root = ../paper.tex
%-------------------------------------------------------------------------------
\section{\prot{} Details}
\label{sec:protocol}
%-------------------------------------------------------------------------------
\begin{figure}[t]
    \centering
    \begin{tcolorbox}[colback=white!10,
                      width=\columnwidth,
                      arc=1mm, auto outer arc,
                      boxrule=0.3pt, 
                      left=0pt,
                     ] 
    \begin{tabular}{r@{\hspace{0.1cm}} >{\raggedright\arraybackslash}p{6.3cm}}
        $\mathit{etaQ}$ :& Priority queue of requests, ordered by ETA. \\
        $\mathit{log}$ :& Log of the replica. \\
        $\mathit{chkpt}$ :& Latest committed checkpoint. \\
        $i$ :& Round number for \repair{}, incremented after every invocation of \repair{}.\\
        $v$ :& Internal view number for \repair{} subprotocol, which determines the leader. The notion of a leader is only relevant to \repair{}.\\
    \end{tabular}

    \end{tcolorbox}
    \caption{Replica state in \prot{}.}
    \label{fig:replica_state}
\end{figure}

\subsection{Replica State}

The key state at each replica is summarized in Figure~\ref{fig:replica_state}. 
Replicas maintain the partially speculative $\mathit{log}$. An entry at index $k$ in the log contains the full client request, the execution result, and a hash digest of the log up to $k$, which we will refer to as $H(k)$. We compute $H(k)$ by hashing the request at $k$ and $H(k-1)$. The hash digest allows replicas and clients to quickly check whether replica logs are consistent up to an index $k$. To track the committed prefix of the log, replicas also record the latest checkpoint $\mathit{chkpt}$, which consists of an index in the log ($\mathit{chkpt.idx}$) and a proof of correctness of the checkpoint ($\mathit{chkpt.proof}$). The \syncp{} protocol ensures that all indices up to and including $\mathit{chkpt.idx}$ are committed. 

Replicas must also track the round number $i$, which separates rounds of the fast path from rounds of \repair{}. Each time the \repair{} subprotocol succeeds in agreeing on a common log, replicas increment their round number to  $i + 1$ and return to the fast path, ignoring any messages from previous rounds.

Finally, since \repair{} is a leader-based subprotocol derived from PBFT, it also uses a view number $v$ that determines the leader. However, $v$ and leaders are \emph{internal} to \repair{}; only the messages and replica state within \repair{} use the view number $v$.

\subsection{Setting and Waiting for ETAs}\label{sec:protocol:etas}

To submit a request, clients multicast messages of the form $m_c = \signed{\texttt{REQUEST}, c, s_c, \eta, \mathit{op}}{c}$ to all replicas. Here, $\mathit{op}$ is the operation requested by the client, $c$ is the client identifier, $s_c$ is the sequence number at the client and $\eta$ is the ETA. A request sent by a correct client will have the same ETA at every replica.

To set $\eta$, the client adds a dynamically computed offset $\omega$ to its current clock value when generating a \texttt{REQUEST} message. $\omega$ is computed  using periodic probes that measure message delays between the client and replicas and is intended to be an estimate for the maximum amount of time it takes for a request to reach all replicas. Specifically, the client sends probes tagged with local timestamps, and replicas reply with the receive timestamp, from which a one-way delay sample is computed. Each client maintains a sliding window of recent samples for each replica $r$ and computes a $q$th-percentile estimate over the window, denoted $D_q^r$. Our implementation uses $q = 95$. $\omega$ is then set to $\gamma \cdot \max_r\left[{D_q^r}\right]$, i.e., the maximum $q$th percentile across replicas, multiplied by a heuristic coefficient $\gamma \geq 1$. $\gamma$ can be tuned to make ETAs more conservative or aggressive, essentially trading lower latency against a higher risk of divergence.

\begin{figure}[!t]
    \centering
    \includegraphics[width=\columnwidth]{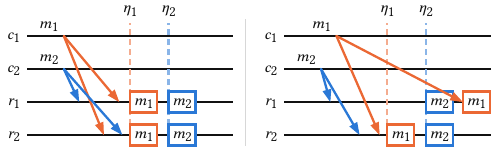}
    \caption{Replicas executing requests by ETA. Left: requests arrive
      on time, so both replicas execute $m_1$ then $m_2$. Right: $m_1$ reaches
      $r_1$ late, and the logs diverge.}
    \label{fig:etas}
\end{figure}

On the replica side, incoming requests are placed into $\mathit{etaQ}$, a priority queue ordered by ETA, with ties 
broken by client ID. When in the fast path, a replica releases the request at the head of $\mathit{etaQ}$ once its
local clock passes that request's ETA. A request that arrives after its ETA
is released immediately, since the local clock has already passed it, though processing it may
cause the replica's log to diverge.
Figure~\ref{fig:etas} illustrates both outcomes: requests that arrive before their ETA are
executed in ETA order at every replica, whatever order they arrived in, while one that
arrives late is executed out of order at the replica that missed it.

Replicas must also cap how far in the future a request may wait. If a request arrives at local time $t$ with an ETA larger than $t+\mathit{maxWait}$, it is queued with its ETA set to $t+\mathit{maxWait}$ instead. This preserves liveness under large clock synchronization errors, since otherwise a request with a far-future ETA could wait arbitrarily long before release. However, this comes at the cost of forcing requests off the fast path.

% \subsection{Fast Path}
To process a request $m_c$ once it has left $\mathit{etaQ}$, 
a replica applies $\mathit{op}$ to the state machine and obtains an execution 
result $\mathit{res}$. The replica then appends the request to its log at the 
next available index $k$, and computes $H(k)$ by hashing the message $m_c$ and the previous 
log entry's digest $H(k-1)$. Then it 
replies directly to the client with a $\signed{\texttt{SPEC-REPLY}, i, c, s_c, k, H(k), \mathit{res}}
{r}$ message. If the client gathers $(n - p)$ consistent (i.e., equal in all fields 
except replica ID) \texttt{SPEC-REPLY} messages, it can deliver $\mathit{res}$ to the underlying 
application as committed in the \textbf{fast path}.

\subsection{Sync Subprotocol}
\label{sec:checkpointing}

The \syncp{} subprotocol creates checkpoints to limit the size of the uncommitted portion of $\mathit{log}$ and allows replicas to detect divergence. Appendix~\ref{sec:appendix:pseudo} gives its pseudo-code.

When its log reaches a multiple of $I$ entries, a replica $r$ broadcasts $\signed{\texttt{SYNC}, i, k, H(k), \eta_k^*}{r}$, where $i$ is its round number, $k$ is the index of its last log entry, $H(k)$ is the log digest up to $k$, and $\eta_k^*$ is the largest ETA of any request in $r$'s log up to $k$, which \alignp{} uses to decide which requests to replay after resetting a replica's state (\S\ref{sec:protocol:align}).

If replicas do not reach a multiple of $I$, they also send \texttt{SYNC} messages based on a timer for $k = |\mathit{log}| - 1$. Since logs may then differ in length, a replica that receives a \texttt{SYNC} for an index $k' < |\mathit{log}|$ it has not sent one for broadcasts its own \texttt{SYNC} for $k'$.

Effectively, these entry points to \syncp{} may cause replicas to coordinate on multiple
indices in parallel. For any such index $k$, each replica can either successfully complete 
a checkpoint as part of the fast path quorum, realize it needs to \alignp{} to an existing 
quorum, or fall back to \repair{} to fix its log if no such quorum exists.

\Para{Checkpoint.}
Once a replica $r$ gathers $(n - p)$ consistent \texttt{SYNC} messages for an index $k$, the log is committed up to and including $k$, since a consistent $H(k)$ implies the earlier requests were also consistent. $r$ then updates $\mathit{chkpt}$ if $k >\mathit{chkpt.idx}$, storing the $(n - p)$ \texttt{SYNC} messages as $\mathit{chkpt.proof}$, and broadcasts $\signed{\texttt{CHECKPOINT}, i, k, H(k)}{r}$.

\Para{Align.}
If a replica sees $f + 1$ consistent \texttt{CHECKPOINT} messages that conflict with its own state, it knows it has diverged, and triggers the \alignp{} subprotocol. \alignp{} runs in the background and is off the fast path's critical path.

\Para{Repair.}
Finally, if there is not a large enough quorum to create a checkpoint, the replicas need to invoke the \repair{} subprotocol. Replicas decide to enter \repair{} by either timing out or gathering a conflict proof (\S\ref{sec:repair:trigger}).

\subsection{Alignment Subprotocol} \label{sec:protocol:align}

\begin{figure}[!tbp]
    \centering
    \includegraphics[width=\columnwidth]{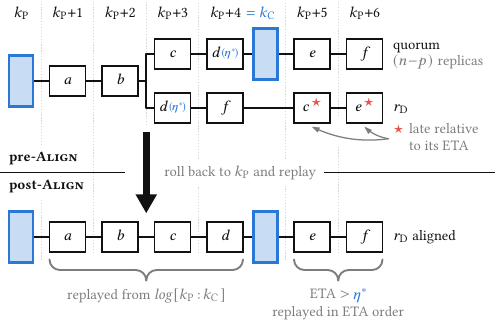}
    \caption{Example of \alignp{}. Requests $a$--$f$ were sent with
    increasing ETA. $\rd$ diverged after receiving $c$ and $e$ late.
    Replaying by ETA order allows $\rd$ to match the quorum.}
    \label{fig:align_logs}
\end{figure}

The \alignp{} subprotocol aims to align the log of a diverged replica $\rd$ to the common log of the replicas that created a checkpoint. \alignp{} involves not only a state transfer for a checkpoint, but also requires $\rd$ to infer the order of any post-checkpoint requests it has already received, and then speculatively re-execute these requests after resetting to the checkpoint.

When $\rd$ enters \alignp{} (after seeing $f + 1$ consistent \texttt{CHECKPOINT} messages for index $\kc$), it stops releasing requests from $\mathit{etaQ}$ and sends a $\signed{\texttt{STATE-REQUEST}, \kc, \kp}{}$ to all of the replicas that have sent \texttt{CHECKPOINT} messages. Here $\kp$ is set to $\mathit{chkpt.idx}$, the index of the last checkpoint $\rd$ committed, or $0$ if no such checkpoint exists. When a replica $r$ receives the \texttt{STATE-REQUEST}, if it has the checkpoint, it replies with a $\signed{\texttt{STATE-REPLY}, \kc', \mathit{log}[\kp:\kc'], \mathcal{S}}{r}$. Here, $\kc'$ is the index of the latest checkpoint at $r$, which may be larger than the requested index $\kc$. $\mathcal{S}$ is the set of $n - p$ \texttt{SYNC} messages that prove the validity of the state at $\kc'$, and $\mathit{log}[\kp:\kc']$ is the portion of the log from $\kp$ to $\kc'$ that can be used by $\rd$ to reset its state to match the quorum at $\kc'$.

Figure~\ref{fig:align_logs} follows a diverged replica through the rest of \alignp{}. $\rd$ rolls back its state to $\kp$ and replays the requests from $\mathit{log}[\kp:\kc']$, bringing its state equal to the state the checkpoint at $\kc'$ represents. $\rd$ must then decide which requests to process next. This is challenging because $\rd$ may have processed requests at slots higher than $\kc'$ before requesting alignment, and has no knowledge of the order the fast path quorum used past $\kc'$. The solution is surprisingly simple, and exploits the fact that the quorum is most likely processing requests in ETA order.
Specifically, $\rd$ reads $\eta^*$, the largest ETA of any request in the log up to $\kc'$, from the \texttt{SYNC} messages in $\mathcal{S}$. It then merges each request from its old, rolled back log whose ETA is greater than $\eta^*$ into $\mathit{etaQ}$. Since every request in $\mathit{log}[\kp:\kc']$ has an ETA of at most $\eta^*$, this naturally eliminates duplicates, and allows $\rd$ to transition smoothly back to the fast path, releasing requests in ETA order as its clock passes them. If after aligning $\rd$ receives all requests  earlier than their ETAs, and the quorum also executes in ETA order, $\rd$ will be able to rejoin the fast path.

Note that a request in the previous log of $\rd$ may have an ETA less than $\eta^*$ and still not appear in $\mathit{log}[\kp:\kc']$. This can occur if the fast path quorum receives a request late after the checkpoint was created. Rather than discard these requests, \prot{} retains them in an orphan set $\mathcal{O}$ for potential reprocessing in a subsequent \repair{}.

As part of this processing, $\rd$ sends new $\texttt{SPEC-REPLY}$ messages, which may seem dangerous at first since multiple conflicting $\texttt{SPEC-REPLY}$ messages would be sent for the same index or client request. However, \alignp{} remains safe because any old \texttt{SPEC-REPLY} messages (i.e., from before $\rd$ entered \alignp{}) with index greater than $\kc'$ are implicitly invalid, since they are based on a log that conflicts with the checkpoint at $\kc'$. Thus, clients will only be able to gather at most $p + f$ $\texttt{SPEC-REPLY}$ messages with digests matching a pre-alignment \texttt{SPEC-REPLY} message, $p$ from the diverged replicas and $f$ from faulty replicas. 

\subsection{Repair Subprotocol}\label{sec:protocol:repair}

\begin{figure}[!tbp]
    \centering
    \includegraphics[width=\columnwidth]{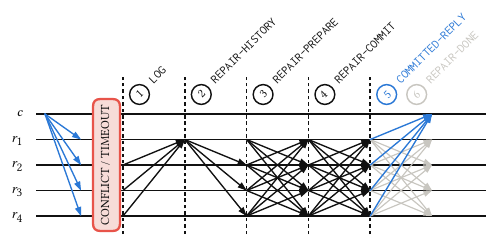}
    \caption{Message flow of the \repair{} subprotocol, numbered as in
    \S\ref{sec:protocol:repair}.}
    \label{fig:repair_messages}
\end{figure}

The \repair{} subprotocol (Figure~\ref{fig:repair_messages}) is a leader-based protocol based on PBFT. At a high level, each invocation of \repair{} is a separate consensus instance whose value is a set of uncommitted replica logs. The uncommitted logs are used to transition from round $i$ to round $i + 1$ and bring \prot{} back into the fast path. 
%In this framing, replicas act as both the clients and replicas of PBFT.
This allows replicas to not only ensure that all committed requests are retained, but to also make progress committing requests if the fast path is unavailable (e.g., due to high network unpredictability or faults). 

The log-gathering phase of \repair{} resembles PBFT's view-change procedure because both construct a new log while preserving previously committed requests. Unlike a PBFT view change, however, \repair{} itself reaches agreement to make progress. This should also be distinguished from \repair{}'s internal view changes, which are used only to replace a faulty or stalled \repair{} leader and ensure liveness.

\subsubsection{Entering Repair}{\label{sec:repair:trigger}}

The first way replicas can enter \repair{} is by timing out while attempting to create a checkpoint during \syncp{}. For a \syncp{} at index $k$, the timer begins once a replica gathers $(n-f)$ \textbf{total} (not necessarily consistent) \texttt{SYNC} messages. The replica cancels the timeout if it creates a checkpoint or observes $(f+1)$ \texttt{CHECKPOINT} messages from other replicas.

If the timer expires, $r$ broadcasts a $\signed{\texttt{TIMEOUT}, i, k}{r}$ message but does not yet enter \repair{}; it waits to see whether other replicas also time out. Once $r$ collects $f+1$ \texttt{TIMEOUT} messages, it combines them into a \texttt{TIMEOUT-PROOF}, broadcasts it, and enters \repair{}. Any replica that receives the \texttt{TIMEOUT-PROOF} likewise broadcasts it and enters \repair{}. This ensures that if one correct replica enters \repair{}, the others soon follow, while requiring at least one correct replica to have actually timed out.

Alternatively, replicas can enter \repair{} as soon as they can prove that a checkpoint is impossible, without waiting for a timeout. Suppose replica $r$ has received $m$ \texttt{SYNC} messages for index $k$, and let $c$ be the size of the largest mutually consistent subset. A checkpoint requires $(n-p)$ consistent \texttt{SYNC} messages. Therefore, if
$n-m < (n-p)-c$,
then even if every remaining \texttt{SYNC} message were consistent with the largest existing subset, $r$ still could not form a checkpoint. The $m$ messages therefore constitute a \texttt{CONFLICT-PROOF}, which $r$ broadcasts; any replica that receives the proof rebroadcasts it and enters \repair{}.

In particular, once a replica has received \texttt{SYNC} messages from all $n$ replicas, there are only two possibilities: either some $(n-p)$ messages are consistent and it can create a checkpoint, or they are not, in which case the $n$ messages themselves form a \texttt{CONFLICT-PROOF}. Thus, when all replicas respond, replicas can normally decide between checkpointing and entering \repair{} without waiting for the timeout.

\subsubsection{Agreeing on History} \label{sec:protocol:repair:pbft}

Once a replica decides to enter \repair{}, it cancels any timeouts, stops accepting \texttt{SYNC} messages, and pauses releasing client requests from $\mathit{etaQ}$. As shown in Figure~\ref{fig:repair_messages}, each replica $r$ begins \repair{} by sending a $\signed{\texttt{LOG}, i, v, \mathcal{L}, \mathcal{O}}{r}$ to the leader of the current view, whose replica ID is $r^* = v \bmod n$~\circled{1}. $\mathcal{L}$ is the uncommitted log at $r$, starting from $\mathit{chkpt.idx}$, with each entry at index $k$ consisting of the log digest $H(k)$ and a client request identifier $(c, s_c, h(\mathit{op}))$, where $h(\mathit{op})$ is a digest of the client operation. $\mathcal{O}$ is $r$'s orphan set (\S\ref{sec:protocol:align}) of client requests that may have been discarded during \alignp{}.

After the leader receives a quorum of $(n-f)$ valid \texttt{LOG} messages for round $i$, it
creates a $\signed{\texttt{REPAIR-HISTORY}, i, v, \mathcal{H}}{r^*}$ and broadcasts it to other replicas~\circled{2}. Here $\mathcal{H}$ is the set of $(n - f)$ \texttt{LOG} messages, or \textit{history}.

On receiving the leader's proposal, all correct replicas run two phases of agreement: in the \textbf{prepare phase}~\circled{3}, replicas broadcast a $\signed{\texttt{REPAIR-PREPARE}, i, v, h(\mathcal{H})}{r}$ message, then attempt to collect a prepare certificate of $(n - f)$ matching \texttt{REPAIR-PREPARE} messages. Here, $h(\mathcal{H})$ denotes the digest of the proposed history. Then, in the \textbf{commit phase}~\circled{4}, each replica sends a $\signed{\texttt{REPAIR-COMMIT}, i, v, h(\mathcal{H})}{r}$ message. Once it gathers $(n - f)$ such messages, the replica may apply $\mathcal{H}$ to deterministically compute a repaired log and enter round $i + 1$. 

Once both phases have completed, replicas broadcast an extra $\signed{\texttt{REPAIR-DONE}, i, v, k, h(\mathcal{H})}{r}$ message after modifying their log and entering the next round~\circled{6}. These messages ensure that if enough correct replicas exit \repair{}, the other correct replicas can exit as well. Specifically, if a replica $r$ sees $f + 1$ \texttt{REPAIR-DONE} messages from distinct replicas, $r$ can safely enter the new round by using the corresponding \texttt{REPAIR-HISTORY} message.

\subsubsection{Constructing the Repaired Log}\label{sec:repair:construct}
\begin{figure}[!tbp]
    \centering
    \includegraphics[width=\columnwidth]{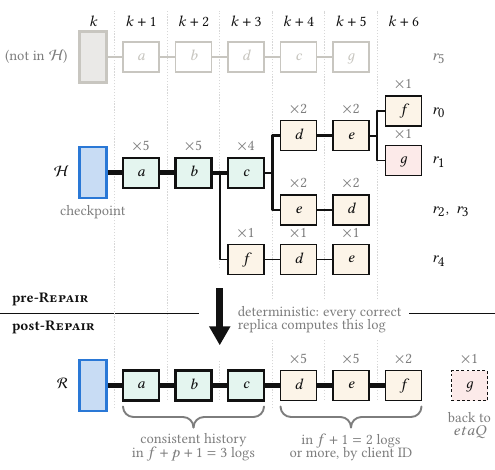}
    \caption{Building the repaired log $\mathcal{R}$ from history $\mathcal{H}$.}
    \label{fig:repair_history}
\end{figure}

The algorithm replicas use to construct a repaired log $\mathcal{R}$ from $\mathcal{H}$
(Figure~\ref{fig:repair_history}) must satisfy two
requirements. First, log construction must be deterministic, so that all correct replicas 
arrive at the same log. Second, the log construction must preserve any requests committed 
in the fast path or in previous rounds of repair.

Replicas can satisfy the second requirement because of the quorum
intersection between the $(n -f)$ logs in $\mathcal{H}$ and the $(n - p)$ replicas
that sent \texttt{SPEC-REPLY} messages for a request committed in the fast path.
Specifically, the $(n-f)$ quorum must intersect the $(n -  p)$ fast path quorum at
$(n - f) + (n-p) - n = 2f + p + 1$ replicas, $f + p + 1$ of which must be correct. 
These correct replicas constitute a majority of the $n - f = 2p + 2f + 1$ replicas 
whose logs are contained in $\mathcal{H}$. Thus, $\mathcal{R}$ keeps any request that
appears with a consistent history (i.e., at the same index and with the same log digest) in
$f+p+1$ logs in $\mathcal{H}$ at that index. Any remaining, non-duplicate requests that
appear in a log or an orphan set at least $f + 1$ times (ensuring they are not fabricated
by malicious replicas) are then appended to $\mathcal{R}$ in a deterministic order, such as
sorted by client ID. All other requests return to $\mathit{etaQ}$.

In Figure~\ref{fig:repair_history}, requests $a$, $b$, and $c$ have consistent histories in $f + p + 1=3$ logs. $d$ and $e$ occupy the same index in $f + p + 1$ logs, but those logs have inconsistent prefixes and therefore digests. But $d$, $e$, and $f$ exist in at least $f + 1=2$ logs, so they are added to $\mathcal{R}$. Finally, any remaining requests like $g$ will not be committed in \repair{}. Note that the contents of $\mathcal{R}$ depend on the choice of logs in $\mathcal{H}$. For example, if $r_5$ were also included in $\mathcal{H}$, $g$ would have appeared $f+1$ times and been included in $\mathcal{R}$.

Logs in $\mathcal{H}$ may start at different checkpoints, for example if some replicas
complete a checkpoint before entering repair and others do not. Since a checkpoint also
requires a fast path quorum of $(n - p)$ replicas, a majority of the logs in $\mathcal{H}$
are consistent with any correct replica's checkpoint, so replicas simply compute
$\mathcal{R}$ from the \emph{highest} valid checkpoint in $\mathcal{H}$. 
\subsubsection{Exiting Repair}

To reset its state to the agreed upon repaired log, a replica $r$ finds the first index in the new log that does not match its current log, then rolls back its state to that index and reapplies any requests from the repaired log. Any requests that are not in the repaired log (such as $g$ on $r_1$ in Figure~\ref{fig:repair_history}) will be reinserted into $\mathit{etaQ}$ to be retried in the next round. Since the logs in $\mathcal{H}$ only contain metadata about client requests, $r$ may need to fetch some client request payloads from other replicas, which it can do by simply retrieving them from any of the $f+ 1$ replicas that included a request in the log.

For every request in the newly committed repaired log, $r$ sends a $\signed{\texttt{COMMITTED-REPLY}, i, c, s_c, k, H(k), \mathit{res}}{r}$ to the client $c$ that sent the request~\circled{5}, where $k$ is the index of the request in the repaired log. Since the request is committed at the replica, the client only needs to wait for $f + 1$ \texttt{COMMITTED-REPLY} messages to safely deliver the result.

The replica $r$ can also create a new checkpoint at the last index of the repaired log, using the $(n - f)$ \texttt{REPAIR-COMMIT} messages as a proof of correctness, rather than $(n - p)$ \texttt{SYNC} messages. Thus, in the subsequent round of \repair{}, replicas will not need to include any requests committed in the previous round. This allows \repair{} to make consistent progress even in the face of consecutive repair rounds.

Finally, to exit \repair{} and enter the next round: $r$ increments its round number to $i + 1$ and returns to the fast path. It begins releasing requests from $\mathit{etaQ}$ again, with any requests that were committed during repair removed.

\subsubsection{Internal View Change}

View change in \repair{} is a simplified version of the PBFT view change, and serves the same purpose of replacing faulty or stalled \repair{} leaders. Because only one \repair{} instance is active at a time, replicas need to include only a single prepare certificate for the current round rather than a full log.

% Replicas also  proactively send their original \texttt{LOG} message alongside their \texttt{VIEW-CHANGE} messages to the new leader, which allows the new leader to immediately propose a new \texttt{REPAIR-HISTORY} if none was prepared in the previous view.

The main complication is the interaction between \repair{} views and \prot{} rounds. Replicas start their view-change timers when they enter \repair{}. Dissemination of the \texttt{TIMEOUT-PROOF} or \texttt{CONFLICT-PROOF} ensures that correct replicas enter \repair{} at roughly the same time. Therefore, under a stable network, a correct leader should then make progress before these timers expire. If not, replicas initiate a view change. A replica that has already completed \repair{} and entered round $i+1$ must still participate in view changes for round $i$ so that lagging replicas are not left behind. Conversely, if a replica has begun a view change in round $i$ but later receives $f+1$ \texttt{REPAIR-DONE} messages for that round, it adopts the corresponding \texttt{REPAIR-HISTORY} and enters round $i+1$. If round $i+1$ later enters \repair{}, the replica resumes advocating the pending view change.

\subsection{Practical Considerations}\label{sec:practical}

\begin{outline}
\Para{Snapshotting and Garbage Collection.}
As described, replicas can never truncate their logs, since other replicas may need them for \alignp{} and \repair{}. As in other SMR protocols, \prot{} can periodically take application snapshots, include the snapshot digests in \texttt{SYNC} and \texttt{CHECKPOINT} messages, and truncate the log once a snapshot is attested to by \texttt{CHECKPOINT} messages from $f+1$ replicas. \alignp{} and \repair{} may then transfer snapshots in place of log segments.

\Para{Queuing Requests.}
Requests that arrive during \repair{} and \alignp{} are queued in $\mathit{etaQ}$ rather than discarded. Since no requests are released during \repair{}, requests that arrive late relative to their ETAs are still ordered correctly, as long as they arrive before \repair{} finishes. If clients are correct, these requests then commit on the fast path immediately after \repair{}, letting \prot{} quickly drain the queue of requests caused by \repair{}'s disruption of the fast path (\S\ref{sec:eval:repair}).

\Para{Garbage Collecting Queued Requests.}
Since \repair{} returns any request that does not commit (i.e., one that appears in fewer than $f + 1$ replica logs) to $\mathit{etaQ}$, a request from a crashed client that reached only some replicas could be retried indefinitely. After each \repair{}, replicas that retained the request may execute it again while others do not, which would immediately cause their logs to diverge and trigger another \repair{}. To prevent this, replicas expire queued requests after a fixed $\mathit{TTL}$, refusing to re-queue them after repair.
\end{outline}

\section{Evaluation}\label{sec:evaluation}
%-------------------------------------------------------------------------------
Our evaluation seeks to answer the following questions:

\begin{enumerate}[nosep]

    \item \textbf{Performance: } How does \prot{} compare to other state-of-the-art protocols in terms of throughput and latency in the normal, fault-free case? (\S\ref{sec:eval:baselines})

    \item \textbf{Latency: } How does \prot{}'s best case latency of $2\Delta + \epsilon$ translate to latency improvements across deployments? (\S\ref{sec:eval:latency})

    \item \textbf{ETA Estimation: } How does the ETA estimation affect \prot{}'s ability to stay in the fast path? (\S\ref{sec:eval:eta})
    
    \item \textbf{Recovery: } How does \repair{} affect the performance of \prot{}? (\S\ref{sec:eval:repair})

    \item \textbf{Extra Replicas and Alignment: } How much do extra replicas and \alignp{} contribute to \prot{} staying in the fast path? (\S\ref{sec:eval:extra})

    \item \textbf{Scaling: } How well does \prot{} scale as the total number of replicas grows? (\S\ref{sec:eval:scaling})
\end{enumerate}

\subsection{Setup}

\Para{\prot{} Implementation.} \prot{} is implemented in about 20{,}000 lines of Rust. 
Our implementation uses TCP for communication, the \texttt{ed25519-dalek} crate for digital signatures and the \texttt{sha2} crate for SHA-256 digests. 

\Para{Testbed.} We use \texttt{n4-standard-16} VMs in 12 North American Google Cloud regions for all of our experiments. All 12 regions host clients, and 6 of them also host replicas. More details about the testbed are presented in Appendix~\ref{sec:appendix:testbed_measurements}. VM clocks are synchronized with Google Cloud's default NTP setup \cite{chrony}, leading to sub-millisecond clock error.

\Para{Baselines.}
We compare \prot{} against several existing low-latency BFT protocols.  PBFT~\cite{pbft} and Zyzzyva~\cite{zyzzyva} are well-known leader-based protocols, with Zyzzyva improving on the latency of PBFT by adding a speculative, leader-based $3\Delta$ fast path. Flutter~\cite{flutterblink} is a leaderless protocol that, like \prot{}, uses delay estimates, but does not speculate and therefore has a $3\Delta$ fast path. Autobahn~\cite{autobahn} is a high-throughput consensus protocol %built around a \textit{mempool} 
that separates request dissemination from consensus, while also attempting to reduce latency through an optimistic fast path. Banyan~\cite{banyan} is a rotating-leader consensus protocol that uses fast-path quorums in parallel with normal consensus. Table~\ref{tab:protocol_message_delays} summarizes the theoretical best-case latency and optimistic techniques used by each protocol.

We implement PBFT, Zyzzyva, and Flutter in our Rust framework. PBFT and Zyzzyva use leader-based batching, with a batch size of 20 to improve throughput while adding minimal batching delay\ifdefined\fullappendix{} (see Appendix~\ref{sec:appendix:variants:batch} for comparisons)\fi. Flutter does not prescribe a specific delay-estimation mechanism, so we use the same ETA estimator as \prot{}. Flutter also requires clients to retry when a request misses its estimate and is rejected; we implement this with exponential backoff and jitter.

Autobahn is also written in Rust and uses the same transport and cryptography libraries as \prot{}, while Banyan is implemented in Go. 
However, unlike most BFT evaluation setups in the literature~\cite{pbft,zyzzyva, HQosdi, aardvark}, 
these implementations do not have client request authentication. 
%are unable to be compared fairly to \prot{} and our baselines as published, since they defer client request verification to the application rather than handling it as part of the protocol. 
To enable apples-to-apple comparison, we therefore implement a client protocol on top of each, in which clients sign requests and verify replies, and replicas verify each request\footnote{A similar modification was made in the evaluation of Chop Chop~\cite{chopchop} when comparing against Narwhal~\cite{narwahl_tusk}, which Autobahn's implementation forks.}, and call the resulting protocols \textbf{Autobahn-RSM} (Replicated State Machine) and \textbf{Banyan-RSM}. Compared to the published evaluations, the RSM variants add latency to the client-replica request and reply paths, since clients are no longer co-located with a replica. For Autobahn-RSM, per-client-request authentication also affects its throughput.  By contrast, Banyan-RSM is  limited by leader rotation, so client authentication does not affect its throughput  (\S\ref{sec:eval:baselines}). %Banyan-RSM is instead limited by leader rotation, so the client protocol does not affect its throughput.

\subsection{Performance Relative to Baselines} \label{sec:eval:baselines}

\begin{figure}[!t]
  \centering
  \includegraphics[width=\columnwidth]{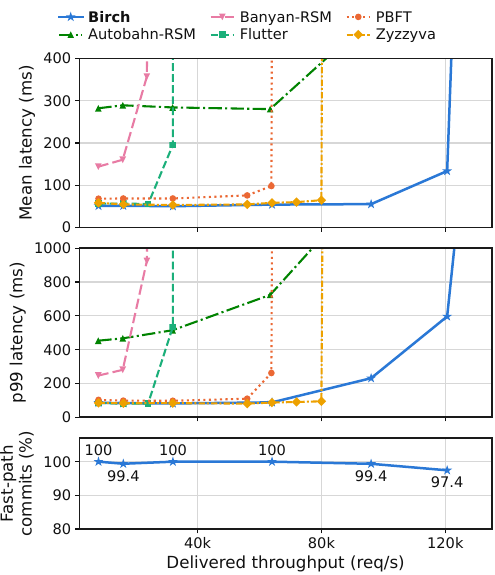}
  \caption{\prot{} has the lowest mean latency and highest peak throughput of the protocols we evaluate, and maintains high fast path utilization (\SI{1}{\kibi\byte}, $n=6$, $f=1$).}
  \label{fig:latency_vs_throughput_baselines}
\end{figure}

We first evaluate each protocol by measuring both latency and throughput. For each protocol, we increase the offered load and report latency and throughput up to the point where the protocol saturates. Clients are open loop and send \SI{1}{\kibi\byte} payloads. Figure~\ref{fig:latency_vs_throughput_baselines} shows mean and 99th percentile latency against delivered throughput, as well as the fraction of requests \prot{} commits on the fast path for each trial.\ifdefined\fullappendix{} Appendix~\ref{sec:appendix:variants:16b} also repeats this comparison with \SI{16}{\byte} payloads.\fi

The key result is that \prot{} commits requests with the lowest latency while matching or exceeding the throughput of every baseline. Compared to the leader-based protocols, \prot{} improves mean latency under light load by 7--\SI{17}{\milli\second} and sustains $1.6\times$--$2.0\times$ their throughput. Compared to Autobahn-RSM, it commits in less than a fifth of the time and delivers $1.3\times$ the throughput in our chosen scenario.

In more detail, \prot{} has a mean latency of \SI{51}{\milli\second} under light load, and it reaches a peak throughput of 129k requests/s. \prot{} commits at least $99\%$ of requests on the fast path up to 96k requests/s, and $97\%$ at 120k.\ifdefined\fullappendix{} The fast path also holds over time: over a one hour run at 50k requests/s, $99.84\%$ of commits take it (Appendix~\ref{sec:appendix:longrun}).\fi{} Although slow-path commits have little effect on throughput or mean latency, they are visible in the tail latency, as seen by \prot{}'s p99 latency rising before saturation.

The fast path gives \prot{} both low latency and efficient request processing: clients send requests directly to every replica, so no single leader carries the full load. Autobahn-RSM similarly avoids a single leader and peaks at 103k requests/s, but with a much higher mean latency of \SI{282}{\milli\second} under light load. Both systems are bottlenecked by client-request cryptography rather than consensus. PBFT and Zyzzyva, by contrast, must broadcast every request through a leader and saturate at 64k and 80k requests/s, respectively. Under light load, Zyzzyva's mean latency is \SI{58}{\milli\second} and PBFT's is \SI{68}{\milli\second}. The remaining baselines saturate much earlier. Banyan-RSM is limited by its rotating leader and peaks at 24k requests/s, with a mean latency of \SI{144}{\milli\second} under light load. Flutter has a mean latency of \SI{56}{\milli\second} under light load, but enters a retry storm once requests begin missing their deadlines above 32k requests/s.

\subsection{Latency Improvements} \label{sec:eval:latency}

\begin{figure*}[!t]
  \centering
  \begin{subfigure}[t]{0.64\textwidth}
    \centering
    \includegraphics[width=\textwidth]{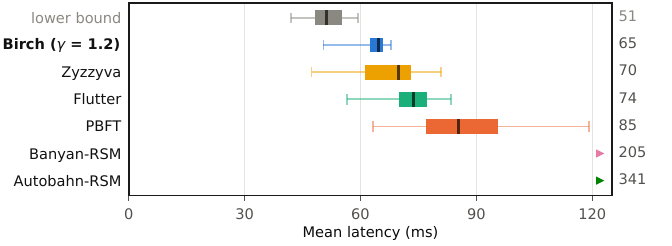}
    \caption{\prot{} shows the best latency across deployments. Mean latency of the median deployment is printed on the right. Autobahn-RSM and Banyan-RSM are off the scale, so only their mean latency is shown.}
    \label{fig:latencysim}
  \end{subfigure}
  \hfill
  \begin{subfigure}[t]{0.34\textwidth}
    \centering
    \includegraphics[width=\textwidth]{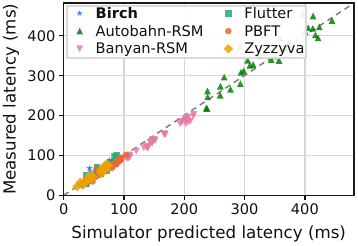}
    \caption{Simulated latencies accurately predict measured median latencies.}
    \label{fig:geo_validate}
  \end{subfigure}
  \caption{Simulator results.}
  \label{fig:geosim}
\end{figure*}

\prot{}'s actual latency savings depend on both how ETAs are set and the replica placement, and therefore cannot be predicted by simply counting message delays. Specifically, \prot{} must wait for ETAs estimated based on the farthest replica and scaled by $\gamma$. %By contrast, although leader-based protocols add an extra hop through the leader, this detour can be cheap, or even faster than a direct path since the triangle inequality does not always hold in the cloud. 
Therefore, we further investigate how replica placements in different deployments can impact protocol latency.  In order to study a large number of diverse deployments, we simulate \prot{} and each baseline's latency based on measured message delays. 

We measure pings between all pairs of the 12 North American Google Cloud regions (600 samples over 10 minutes per link, Appendix~\ref{sec:appendix:testbed_measurements}) and use each link's median as its delay. We simulate every placement of $n=6$ replicas in distinct regions, counting each leader position separately. We also include a network ``lower bound'' of one round trip to the $(n-\p)$th (i.e., 5th) closest replica, with no ETA wait. \prot{} and Flutter use $\gamma = 1.2$ and estimate ETAs from the 95th percentile of each link's samples, matching our implementation.

For each deployment, we gather the simulated latencies of clients in each of the 12 regions and record their mean. Figure~\ref{fig:latencysim} shows the distribution of these means across deployments (median, quartiles, min and max). \prot{} is fastest overall, with the lowest median and the narrowest spread. The gap between the lower bound and \prot{} shows the cost of waiting for ETAs set by the farthest replica. Zyzzyva has lower mean latency than \prot{} on some deployments, but only for certain leaders, which cannot be relied on since they are a result of the current view and not chosen by the operator. Autobahn-RSM and Banyan-RSM are much slower on every deployment, paying extra message delays as well as waits for batches or leader rotation. 

Separately, to validate the accuracy of our simulation, we run each protocol on three random deployments at a low rate of 1{,}000 requests/s, with PBFT and Zyzzyva unbatched, and compare the simulated and measured median latency per client. The results are displayed in Figure~\ref{fig:geo_validate} and show that our simulation captures the actual latency well, with an average error of about 10\% or less for all protocols. The error comes from cloud routing changes, message processing that the simulation ignores, and simplified models of Autobahn's mempool and Banyan's leader rotation.

\subsection{Effect of ETAs on \prot{} Performance}\label{sec:eval:eta}

\begin{figure*}[!t]
  \begin{minipage}[t]{0.32\textwidth}
    \centering
    \includegraphics[width=\textwidth]{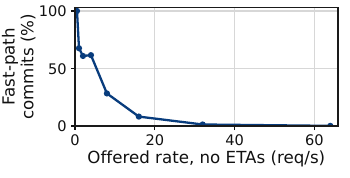}
    \caption{Without ETAs, leaderless fast paths quickly become ineffective.}
    \label{fig:noeta}
  \end{minipage}
  \hfill
  \begin{minipage}[t]{0.66\textwidth}
    \centering
    \includegraphics[width=\textwidth]{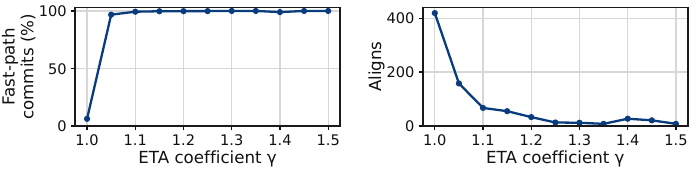}
    \caption{Increasing ETA coefficient $\gamma$ rapidly improves fast path utilization and reduces divergence, as measured by the number of alignments (50k requests/s).}
    \label{fig:gamma_sweep}
  \end{minipage}
\end{figure*}

\prot{} relies on ETA ordering to remain in the fast path. Without the tentative order that the ETAs provide, requests from different clients quickly start to arrive in different orders on replicas, precluding leaderless protocols that do not use ETA ordering (such as Q/U~\cite{QU}, HQ~\cite{HQosdi}, and Aliph~\cite{700bftaliph}) from taking the fast path. To demonstrate this, we run \prot{} with ETA ordering disabled: clients do not estimate ETAs and replicas process requests on arrival. Figure~\ref{fig:noeta} shows that at merely 32 requests/s, fast path utilization reaches 0\%.

We then investigate the effect of clients setting ETAs in more detail. Recall that clients set ETAs based on the maximum 95th percentile delay across all replicas, multiplied by a coefficient $\gamma \geq 1$. Larger values of $\gamma$ result in later ETAs, giving more leeway for requests to arrive on time but costing more waiting time. We illustrate this tradeoff in Figure~\ref{fig:gamma_sweep}. We run \prot{} at 50k requests/s, across values of $\gamma$ from $1$ to $1.5$ for 5-minute trials, and show the fast path utilization and the number of completed runs of \alignp{}. At $\gamma = 1$ the ETA has no headroom beyond the 95th percentile estimate itself and only $6\%$ of requests commit on the fast path. Increasing $\gamma$ to just $1.05$ raises fast path utilization to $97\%$ with 158 runs of \alignp{}, and at $\gamma = 1.1$ utilization exceeds $99\%$ with 67 runs of \alignp{}. Past that point, \repair{} and \alignp{} run only rarely, but random delay spikes still occasionally trigger them. The cost of later ETAs is a slow rise in latency: one client's median latency grows from \SI{37}{\milli\second} at $\gamma = 1.05$ to \SI{46}{\milli\second} at $\gamma = 1.5$.

\subsection{Recovery} \label{sec:eval:repair}

\begin{figure*}[!t]
  \begin{minipage}[t]{0.32\textwidth}
    \centering
    \includegraphics[width=\textwidth]{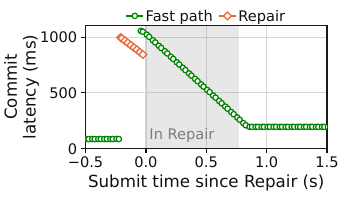}
    \caption{Requests that arrive during \repair{} wait for it to finish, but are committed in the fast path.}
    \label{fig:repair}
  \end{minipage}
  \hfill
  \begin{minipage}[t]{0.32\textwidth}
    \centering
    \includegraphics[width=\textwidth]{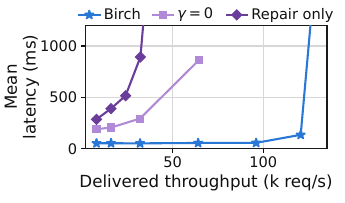}
    \caption{\prot{} performs reasonably well when ETAs are inaccurate and when forced to use \repair{}.}
    \label{fig:adverse}
  \end{minipage}
  \hfill
  \begin{minipage}[t]{0.32\textwidth}
    \centering
    \includegraphics[width=\textwidth]{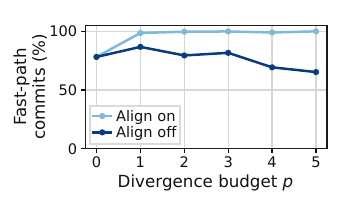}
    \caption{Increasing the number of extra replicas ($2p$) allows more fast path utilization, but only when paired with \alignp{} ($f = 1$, $\gamma = 1.05$).}
    \label{fig:extra_replicas}
  \end{minipage}
\end{figure*}

\prot{} is not always able to take the fast path. \repair{} is slow and expensive; replicas must agree on $O(n)$ uncommitted histories, and cannot process new requests while doing so. In our other experiments, \repair{} does not noticeably reduce throughput, since it only occurs rarely. Here we investigate how \repair{} affects \prot{} in less ideal scenarios.

We first show how a single instance of repair affects latency, and how the queue is drained. Figure~\ref{fig:repair} shows the commit latency of requests against their submission time during a \repair{} round. A small number of requests arrive after replicas diverge but before \repair{} starts; these are committed in the repaired log during \repair{}. Requests submitted during \repair{} are queued and later committed via the fast path when \repair{} completes (\S\ref{sec:practical}). Eventually, \prot{} finishes draining the queued requests on the fast path and returns to processing new requests. Thus, occasional \repair{} events affect the tail latency, but do not affect throughput because the fast path catches up on the requests that were blocked during \repair{}.

What if \repair{} happened more frequently? Figure~\ref{fig:adverse} measures \repair{} under two non-ideal conditions, alongside unmodified \prot{}. First, we set $\gamma = 0$, which emulates a scenario where requests are consistently late, for example due to poor clock synchronization or an unstable network. Some requests queued during \repair{} can still commit on the fast path, but newly arriving requests quickly cause replicas to diverge again. With $\gamma = 0$, \prot{} has a mean latency of \SI{188}{\milli\second} under light load and reaches a peak throughput of 65k requests/s. Second, we force every request to commit in \repair{}, the worst case for \prot{}. Under these conditions, \prot{}'s mean latency rises from \SI{285}{\milli\second} under light load to \SI{2.4}{\second} at its peak of 39k requests/s.

\begin{table*}[!tp]
\small
    \centering
    \setlength{\tabcolsep}{5.5pt}
    \begin{tabular}{lcccccl}
        \toprule
        \textbf{Protocol} & \textbf{Number of Replicas} & \textbf{S} & \textbf{F} & \textbf{L} & \textbf{Best Case Latency} & \textbf{Conditions} \\
        \midrule
        PBFT \cite{pbft} \footnotesize{\color{gray}(1999)} & $3f + 1$ & \xmark & \xmark & \xmark & $5\Delta$ & Correct Leader \\
        \hspace{3pt} {\color{gray} tentative exec.} & $3f + 1$ & \cmark & \xmark & \xmark & $4\Delta$ & Correct Leader \\

        Q/U \cite{QU} \footnotesize{\color{gray}(2005)}   & $5f + 1$ & \cmark & \cmark & \cmark & $2\Delta$ & Correct Clients, No Concurrent Clients  \\
        HQ \cite{HQosdi} \footnotesize{\color{gray}(2006)} & $3f + 1$ & \cmark & \xmark & \cmark & $4\Delta$ & Correct Clients, No Concurrent Clients \\

        Zyzzyva \cite{zyzzyva} \footnotesize{\color{gray}(2007)} & $3f + 1$ & \cmark & \cmark & \xmark & $3\Delta$ & $n$ Correct Replicas \\

        Aliph \cite{700bftaliph} \footnotesize{\color{gray}(2010)} & $3f + 1$ & \cmark & \cmark & \cmark & $2\Delta$ & Correct Clients, No Concurrent Clients \\

        Banyan \cite{banyan} \footnotesize{\color{gray}(2024)} & $3f + 2p^* - 1$ & \xmark & \cmark & \xmark & $4\Delta$ & Correct Leader, $(n-p^*)$ Replicas  \\
        Flutter \cite{flutterblink} \footnotesize{\color{gray}(2024)} & $5f + 1$ & \xmark & \cmark & \cmark & $3\Delta + \epsilon$ & Correct Clients, Delay Estimates \\
        Autobahn \cite{autobahn} \footnotesize{\color{gray}(2024)} & $3f + 1$ & \xmark & \cmark & \xmark & $6\Delta$ & Correct Leader \\

        \textbf{\prot{}} \footnotesize{(this work)} & $3f + 2\p + 1$ & \cmark & \cmark & \cmark & $2\Delta + \epsilon$ & Correct Clients, $(n-p)$ Replicas, Delay Estimates \\
        \bottomrule
    \end{tabular}

\caption{
    Best case end-to-end latency, number of replicas ($p^* \geq 1$, $p \geq 0$), and the fast path conditions of \prot{} and prior SMR protocols.
    Additional columns classify optimistic techniques:
    \textbf{S} (speculative execution),
    \textbf{F} (fast-path quorums),
    and
    \textbf{L} (leaderless).
    % Note that the latencies for Autobahn, Banyan, and Flutter differ from those reported in their papers, as they do not include replica-client message delays.
    }
    \label{tab:protocol_message_delays}
\end{table*}

\subsection{Extra Replicas and Alignment} \label{sec:eval:extra}

\prot{} has a $p$ parameter, which adds $2p$ replicas to the system to allow the fast path to continue despite up to $p$ diverged replicas. However, the benefit of increased replicas depends on how divergence is distributed. For example, if late requests are correlated across the replicas, then replicas would diverge at the same time, making extra replicas useless.

Rather than attempt to characterize the distribution of divergence events, we test the efficacy of extra replicas directly. We run \prot{} with $f=1$ and $\gamma=1.05$ at a fixed request rate of 30k requests/s for 3 minutes, and vary $p$ from $0$ to $5$. We choose a smaller $\gamma$ than in the previous experiments in order to make divergence events more frequent. Figure~\ref{fig:extra_replicas} summarizes the results. The key result is that the fast path utilization rises as soon as there is any divergence budget. Just setting $p=1$ takes the fast path utilization from roughly $78\%$ to near $98\%$, and it stays above $99\%$ as $p$ grows further. 

We also run the same experiment with \alignp{} disabled, which shows that the extra replicas are not as helpful on their own if diverged replicas cannot be aligned. At $p = 0$ any diverging replica causes \repair{}, so \alignp{} is never used and the two configurations coincide. Beyond that point, without \alignp{} a diverged replica never rejoins the quorum, so the extra replicas only delay the onset of \repair{} rather than avoiding it. Past $p = 3$, the delay is outweighed by the cost of \repair{} growing with $n$, and fast path utilization declines.

\subsection{Scaling} \label{sec:eval:scaling}

\begin{table}[!t]
  \centering
  \small
  \begin{tabular}{lcccc}
    \toprule
    $n$ & 6 & 11 & 26 & 51 \\
    \midrule
    Throughput (k req/s) & 129 & 122 & 107 & 99 \\
    Mean latency (ms) & 51 & 65 & 73 & 86 \\
    \bottomrule
  \end{tabular}
  \caption{\prot{} throughput declines gradually as $n$ increases ($f = p$).}
  \label{tab:scaling}
\end{table}

Many modern applications of BFT scale to much larger deployments. Although we did not design \prot{} with scaling specifically in mind, in theory, it should perform well since both the messaging and cryptography costs for the fast path are $O(1)$ per request at each replica. Table~\ref{tab:scaling} shows the peak throughput and mean latency under light load as scale goes up. We scale $f$ and $p$ together with $f = p$, so $n = 5f + 1$. We deploy replicas in the same 6 regions, and add a proportional number of clients in order to ensure client load generation is not the bottleneck. As expected, throughput decays gently rather than collapsing: $n = 51$ still sustains 99k requests/s, about $77\%$ of the 129k that $n = 6$ sustains.

% !TeX root = ../paper.tex
\section{Related Work}

\Para{Optimistic Protocols and Techniques.} \label{sec:related:optimism}
BFT protocols that combine techniques such as fast path quorums, leaderless ordering, and speculative execution can be broadly classified as \textit{optimistic}. 
This design is particularly well suited to controlled deployments such as permissioned blockchains~\cite{hyperledgerfabric, russinovich2019ccf} and services that seek robustness against rare non-crash failures~\cite{driscoll_byzantine_2003}, rather than fully adversarial settings. Table~\ref{tab:protocol_message_delays} summarizes existing optimistic BFT protocols, their fast-path latency, and the conditions required to achieve it. Most prior protocols use only one or two of these techniques and therefore do not achieve a $2\Delta$ best-case end-to-end latency. Q/U~\cite{QU} and Aliph~\cite{700bftaliph} use all three, but cannot handle concurrent requests. Some prior works, including Autobahn, Banyan, and Flutter, report \textit{commit latency}, which is $1\Delta$ shorter than the end-to-end latency presented here.

\Para{BFT Protocol Composition.} Optimistic protocols must have alternative paths to consensus to handle the cases where these fast path conditions are unavailable. One common solution is to compose BFT protocols, creating a fallback based on an existing protocol \cite{optimisticBA, 700bftaliph, HQosdi}. Other protocols run the fast path in parallel with the alternative consensus mechanism to reduce disruptions from the fast path being unavailable \cite{geng2023nezha, banyan, kudzu, sbft}. \prot{} takes the former approach, falling back to a simplified PBFT-style protocol when its fast path fails. 

\Para{Conflict Tracking.}
Many leaderless consensus protocols, including the quorum protocols that inspired \prot{}, do not create a total order of requests \cite{sosp13-epaxos, giridharan2026optimality, pnyxdb, QU, 700bftaliph}. Instead, they require replicas to track conflicts between requests. Non-conflicting requests can then commit independently, while conflicting requests are resolved by tracking and agreeing on their dependencies to form a partial order. While these protocols may take the fast path more or less often depending on the workload, \prot{} creates a total order, and thus its fast path is independent of the workload. One benefit of conflict tracking that \prot{} does not explore is that some requests can continue to take the fast path while others undergo recovery.

\Para{Clock Accelerated Protocols.}
A similar use of clock synchronization to \prot{} has been proposed in a number of crash-fault tolerant (CFT) protocols in order to speed up consensus~\cite{geng2023nezha,domino,epaxostoq,sosp25-tiga,vldb25-k2}. In the BFT world, Flutter~\cite{flutterblink} is a protocol that also implements BFT consensus atop synchronized clocks. Flutter requires inter-replica coordination on the critical path for every commit and defers execution until a safe order is confirmed, thus precluding speculative execution. This conservative design results in higher latency than \prot{}, as shown in our evaluations, and the protocol overhead also results in lower peak throughput. However, unlike in \prot{}, a malicious client in Flutter cannot directly affect whether correct client requests are able to take the fast path or not.

\Para{Other Offloading.}
Instead of using best-effort clock synchronization, many systems offload ordering to separate (trusted) components \cite{nopaxos, specpaxos}. NeoBFT~\cite{sigcomm23-neobft} offloads ordering to the network by using trusted programmable switches to perform Authenticated Ordered Multicast (AOM). uBFT~\cite{asplos23-ubft} employs RDMA-based disaggregated memory to implement a Consistent Tail Broadcast (CTB) primitive. Both mechanisms resemble our ETAs in that they establish an initial order of requests, but that order is trusted rather than tentative.

\section{Conclusion}
We present \prot{}, a new BFT protocol that achieves near-optimal end-to-end latency. To achieve such an aggressive fast path in practice, \prot{} leverages synchronized clocks to order requests by estimated times of arrival, and stays on the fast path despite intermittent network unpredictability.

%-------------------------------------------------------------------------------
\bibliographystyle{plain}
\bibliography{citations}

\clearpage
\appendix

\section{\syncp{} Pseudo-code}
\label{sec:appendix:pseudo}
Algorithm~\ref{algo:ckpt-align-repair} gives the pseudo-code for \syncp{}.
% !TeX root = ../paper.tex

%% Useful macros for pseudo-code
\algdef{SE}[EVENT]{Event}{EndEvent}[1]{\textbf{upon}\ \text{\small #1}\ \algorithmicdo}{\algorithmicend\ \textbf{event}}%
\algtext*{EndEvent}
\algdef{SE}[DOWHILE]{Do}{doWhile}{\algorithmicdo}[1]{\algorithmicwhile\ #1}%
\newcommand{\tsc}[1]{{\textsc{\small #1}}}
\makeatletter
\NewDocumentCommand{\LeftComment}{s m}{%
  \Statex \IfBooleanF{#1}{\hspace*{\ALG@thistlm}}\(\triangleright\) #2}
\makeatother

% \captionsetup[algorithm]{font=footnotesize}

\begin{algorithm}[H]
\footnotesize
\caption{Replica Log Synchronization}
\label{algo:ckpt-align-repair}
\begin{algorithmic}[1]

%============================================================
%   GENERATING SYNC MESSAGES
%============================================================

\Event{$|\mathit{log}| \bmod I = 0$
 \textbf{or} $\mathit{syncTimeout}$ expires} \label{algo-line:chkpt-condition}
    \label{line:send-sync}
    \State $k \leftarrow |\mathit{log}| - 1 ;\quad d \leftarrow H(k);\quad \eta^*_k \leftarrow \textsc{max-eta}(k)$ \label{algo-line:app-state-snapshot}
    % \State $a \leftarrow \textbf{if}\; |\mathit{log}| \bmod I = 0\; \textbf{then}\; h(appState)\; \textbf{else}\; \bot$
    \State \textsc{broadcast} $\signed{\texttt{SYNC}, i, k, d, \eta^*_k}{r}$ \label{algo-line:sync-broadcast}
    \State $\mathit{syncTimeout}$.\textbf{reset}() \label{algo-line:sync-timeout-reset}
\EndEvent

%============================================================
%   PROCESSING SYNC MESSAGES
%============================================================

\Event{\textit{receive} $\signed{\texttt{SYNC}, i, k_s, d, \eta^*}{s}$}
    % \LeftComment{Reply to a SYNC for liveness if not currently triggering}
    \If{$(k_s + 1) \bmod I \neq 0 \;\land\; k_s < |\mathit{log}| \;\land\; k_s \notin \mathit{syncQ}$}
        \label{algo-line:different-index}
        \State $d_r \leftarrow H(k_s)$\label{algo-line:new-sync-start}
        \State \textsc{broadcast} $\signed{\texttt{SYNC}, i, k_s, d_r, \textsc{max-eta}(k_s)}{r}$
    \EndIf \label{algo-line:new-sync-end}
    \State $\mathit{syncQ}[k_s][s] \leftarrow d$
    % \LeftComment{If $n - f$ SYNC messages for $k_s$ have arrived, start checkpoint timer}
    \If{$|\{\mathit{syncQ}[k_s]\}| = n - f$} \label{line:start-chkpt-timeout}
        \State $\mathit{chkptTimeout}[k_s]$.\textbf{start()}  %if not already running
    \EndIf

    % \LeftComment{Check for $n - p$ consistent digests}
    \If{$\exists Q \subseteq \mathit{syncQ}[k_s]:\ |Q| = n - p$ 
        $\wedge$ all digests in $Q$ equal $d$} \label{algo-line:ckpt-condition}
        
        \If{$k_s < |\mathit{log}| \,\land\, H(k_s) = d$} 
            \State \textsc{broadcast} $\signed{\texttt{CHECKPOINT}, i, k_s, d}{r}$  \label{algo-line:broadcast-after-ckpt}
            \State $\mathit{chkptTimeout}[k_s]$.\textbf{stop()}  \label{algo-line-cancel-ckpt-timeout}
        \EndIf
    \EndIf
    \State \textsc{check-conflict-proof}$(k_s)$
\EndEvent

%============================================================
%   PROCESSING CHECKPOINT MESSAGES
%============================================================

\Event{\textit{receive} $\signed{\texttt{CHECKPOINT}, i, k, d}{s}$}
    \State $\mathit{chkptQ}[k][s] \leftarrow d$

    % \LeftComment{If $f+1$ consistent checkpoints exist, adopt or align}
    \If{$\exists Q \subseteq \mathit{chkptQ}[k]:\ |Q| = f+1 \wedge$ all $Q$ share $d$} \label{algo-line:align-start}
        \If{$k \geq |\mathit{log}| \lor H(k) \neq d$} \label{line:enter-align}
           \State \alignp$(k)$
        \Else
            \State \textsc{checkpoint}$(k)$
        \EndIf

        \State $\mathit{chkptTimeout}[k]$.\textbf{stop()}  \label{algo-line-cancel-ckpt-timeout2}

    \EndIf \label{algo-line:align-end}

\EndEvent

%============================================================
%   CHECKPOINT IMPOSSIBILITY (CONFLICT PROOF)
%============================================================

\Function{check-conflict-proof}{$k$}\label{line:conflict-proof}
    \State $m \leftarrow |\mathit{syncQ}[k]|$
    \State $c \leftarrow \max\{|S| : S \subseteq \mathit{syncQ}[k] \,\wedge\,
        \forall m_1,m_2 \in S,\, m_1.d = m_2.d\}$

    \If{$n - m < (n - p) - c$}
        \State $\mathcal{C} \leftarrow \mathit{syncQ}[k]$  
        \State \textsc{broadcast} $\signed{\texttt{CONFLICT-PROOF}, \mathcal{C}}{}$
        \State \textsc{repair}()
    \EndIf
\EndFunction

%============================================================
%   TIMEOUT-BASED REPAIR TRIGGER
%============================================================

\Event{$\mathit{chkptTimeout}[k]$ expires} \label{line:timeout}
    % \LeftComment{Have seen $n-f$ SYNCs for $k$ but no checkpoint formed}
    \State \textsc{broadcast} $\signed{\texttt{TIMEOUT}, i, k}{r}$ \label{algo-line:ckpt-timeout}
\EndEvent
\Event{\textit{receive} $\signed{\texttt{TIMEOUT}, i, k}{s}$}
    \State $\mathit{timeoutQ}[k] \leftarrow \mathit{timeoutQ}[k] \cup \{s\}$

    \If{$|\mathit{timeoutQ}[k]| = f+1$} \label{line:timeout-proof} \label{algo-line:timeout-collection}
        \State \textsc{broadcast} $\signed{\texttt{TIMEOUT-PROOF}, \mathit{timeoutQ}[k]}{}$ \label{algo-line:timeout-broadcast}
        \State \repair() \label{algo-line:timeout-repair}
    \EndIf
\EndEvent

\Event{\textit{receive} $m := \signed{\text{\scriptsize\texttt{TIMEOUT-PROOF}}, \mathcal{T}}{}$ \textbf{or} $\signed{\text{\scriptsize\texttt{CONFLICT-PROOF}}, \mathcal{C}}{}$} \label{algo-line:recv-timeout-start}
    \State \textsc{broadcast} $m$
    \State \repair()
\EndEvent \label{algo-line:recv-timeout-end}

% \Event{\textit{receive} $m:=\signed{\texttt{CONFLICT-PROOF}, \mathcal{C}}{}$}
%     \State Broadcast $m$
%     \State \repair()
% \EndEvent

\end{algorithmic}
\end{algorithm}

% !TeX root = ../../paper.tex
\section{Proofs}
\label{sec:appendix:proof}

Here, we prove the safety and liveness of \prot{} as described in the main paper.

\subsection{Safety}

Standard state machine replication~\cite{schneider_smr_1990} defines safety as a property of correct
replicas: every correct replica executes the same sequence of requests in the same order. However,
\prot{} is a \emph{speculative} protocol, meaning a correct replica could execute a request at slot
$k$, roll back its state, and execute a different request at the same slot. Therefore, we instead
define safety using the externally observable effects of the protocol: the set of requests
\emph{delivered at correct clients}. Recall that a correct client returns an execution result to its
application only after gathering sufficient evidence it has been committed. We require that this
global sequence of committed requests forms a consistent total order.

We write $\langle m, k \rangle$ for the placement of a request $m$ at log index $k$, where $m$ is
identified by its client-sequence pair $(c, s_c)$. A message is \emph{for} $\langle m, k \rangle$
when it carries that identifier and index.

We say that a request $m$ is \emph{committed} at slot $k$ if one of the following conditions holds:
\begin{itemize}
    \item \textbf{Fast path.} A correct client gathers at least $n - p$ \emph{consistent}
        \texttt{SPEC-REPLY} messages for its request $m$ at slot $k$ from distinct replicas, within
        a single round $i$. Two \texttt{SPEC-REPLY} messages are \emph{consistent} when they match
        in every field except for replica ID. In particular, matching log digests at index $k$
        imply that the entire log up to and including $k$ is identical across the two replies,
        since each digest is a running hash of the log prefix.
    \item \textbf{Repair.} A correct client gathers at least $f + 1$ consistent \texttt{COMMITTED-REPLY} messages for $m$ at slot $k$, which means at least one correct replica has applied a repair history whose entry at index $k$ is $m$.
\end{itemize}

\begin{theorem}[Safety]
\label{thm:safety}
The requests committed at correct clients form a unique total order.
\end{theorem}

We prove Theorem~\ref{thm:safety} via the lemmas below.
Lemma~\ref{lem:ckpt-consistent} and Corollary~\ref{cor:diverge-bound} establish that although
\alignp{} allows correct replicas to roll back and send multiple \texttt{SPEC-REPLY} messages for a
given slot within a round, only a limited number of replicas will do so.
Lemmas~\ref{lem:fp-fp}--\ref{lem:rp-rp} then use this result to establish that
any two commits in the same round agree on the same request. Finally, Lemma~\ref{lem:cross-round} extends
the result across rounds. The total-order claim then follows because each committed request
is bound to a unique integer slot.

\subsubsection{Quorum Intersections}

\prot{} uses two quorum sizes. The \emph{fast path quorum} of $(n - \p) = 3f + \p + 1$ replicas is
used both for the \texttt{SPEC-REPLY} messages a client collects and for the
\texttt{SYNC} messages that form checkpoints. The \emph{repair quorum} of
$(n - f) = 2f + 2\p + 1$ replicas backs every phase of \repair{}. Here, we show the properties of how these quorums
overlap, which we cite in the subsequent proofs. Two quorums of sizes $a$ and $b$ overlap on at
least $a + b - n$ replicas, of which at most $f$ are Byzantine.
\begin{center}
\small
\begin{tabular}{@{}llcc@{}}
\toprule
     & Quorums                & Overlap         & Correct in overlap \\
\midrule
(Q1) & fast path, fast path   & $3f + 1$        & $\geq 2f + 1$ \\
(Q2) & fast path, repair      & $2f + \p + 1$   & $\geq f + \p + 1$ \\
(Q3) & repair, repair         & $f + 2\p + 1$   & $\geq 2\p + 1$ \\
\bottomrule
\end{tabular}
\end{center}

\subsubsection{Checkpoint Consistency}

Checkpoints and \alignp{} in \prot{} allow a correct replica to roll back
its log within a round, so before reasoning about fast-path commits we must first establish
that the checkpoints of a round cannot conflict, and bound the number of replicas that can
diverge from a checkpoint. Given two checkpoints $C$ and $C'$ with $C.\mathit{idx} \leq C'.\mathit{idx}$, 
the two checkpoints are \emph{consistent} if the log 
prefix of $C'$ up to $C.\mathit{idx}$ matches the log prefix of $C$.
Note that given that we assume collision-resistant digests, this is equivalent to saying that the digests of the two
logs match at $C.\mathit{idx}$, which is how it is checked in the actual protocol. If the two 
checkpoint indices are equal, this implies the two checkpoints must agree on the same log prefix.
A correct replica has \emph{diverged from} a checkpoint $C$ if its
log has reached index $C.\mathit{idx}$ but its digest does not match $C.\mathit{digest}$.

The intuition is that each checkpoint splits the space of possible histories into two sets: those
consistent with the checkpoint and those inconsistent with it. At most $\p$ correct replicas can
send messages for each branch of inconsistent history, but each branch has a different history, so
these can never combine to create an inconsistent checkpoint.

\begin{lemma}
\label{lem:ckpt-consistent}
Any two checkpoints of a round $i$ are consistent.
\end{lemma}

\begin{proof}
We argue by induction on the order in which the checkpoints of round $i$ are created.

\emph{First checkpoint.} Let $C_1$ be the first non-repair checkpoint (i.e., via \texttt{SYNC} messages) created in 
round $i$. No checkpoint $C_1'$ that is inconsistent with $C_1$ can be created at the same index 
as $C_1$, since by (Q1), that would require $2f + 1$ correct replicas to have sent \texttt{SYNC}
messages for both.

\emph{Second checkpoint.} Consider the second checkpoint $C_2$ created in round $i$. Checkpoint $C_2$ must
be consistent with $C_1$. For sake of contradiction, suppose $C_2$ is inconsistent with $C_1$.
Since at least $n - \p$ replicas sent \texttt{SYNC} messages for $C_1$, at most $\p$ correct
replicas could have sent \texttt{SYNC} messages derived from histories inconsistent with $C_1$ --- a
sender of $C_1$ could only come to hold such a history by aligning to a checkpoint inconsistent with
$C_1$, and by hypothesis none exists.
Thus, at most $\p + f$ \texttt{SYNC} messages could have been sent for $C_2$ that are 
inconsistent with $C_1$, which is fewer than the $n - p$ required for $C_2$ to be created.

We must also show that a checkpoint $C_2'$ inconsistent with $C_2$ cannot exist at the same index as $C_2$. 
Again, by (Q1), there must be at least $2f + 1$ correct replicas that sent \texttt{SYNC} messages for both. 
Consider a replica $r$ that sent \texttt{SYNC} messages for both $C_2$ and $C_2'$.
Without loss of generality, assume it sent a message $m'$ before \alignp{} and a message $m$ after \alignp{}, 
corresponding to $C_2'$ and $C_2$ respectively. $m$ must be consistent with $C_1$, since it was sent after \alignp{}.
More subtly, $m'$ must be inconsistent with $C_1$, since it was sent before \alignp{}. In particular, if $m'$ were 
consistent with $C_1$, then the log leading up to $m'$ at the correct replica would also be consistent
with $C_1$ at $C_1.\mathit{idx}$, and the replica $r$ would not have aligned to $C_1$. But previously we showed
any checkpoint after $C_1$ must be consistent with $C_1$, thus this is a contradiction.

\emph{Inductive step.} Consider a checkpoint $C_j$, $j > 1$. Assume that there is only a single consistent history of checkpoints 
$C_1, C_2, \ldots, C_{j-1}$ created in round $i$ on increasing indices such that $C_j.\mathit{idx} > C_{j-1}.\mathit{idx}$. We must show
that $C_j$ is consistent with $C_{j-1}$ (and thus all previous checkpoints). For sake of contradiction, 
suppose $C_j$ is inconsistent with $C_{j-1}$. At least $n - \p$ replicas must have sent a \texttt{SYNC} 
at  $C_j.\mathit{idx}$ indicating a single, identical log inconsistent with $C_{j-1}$. There are two cases:
\begin{enumerate}[wide=0pt, label=\textit{Case \arabic*.}]
    \item This log is inconsistent with every checkpoint $C_k$ for $k < j$. There must be $n - p$
    replicas that reached $C_j.\mathit{idx}$ with this log. However, then $C_1$ would have never been able to be
    created, since at most $\p$ correct replicas would be able to have a history consistent with $C_1$,
    and so there would be at most $\p + f$ \texttt{SYNC} messages for $C_1$, which is fewer than the 
    $n - p$ required to create $C_1$.

    \item This log is consistent with $C_k$ for some $k < j - 1$, and therefore inconsistent with
    $C_{k+1}$. Then, by the same logic, $C_{k + 1}$ would not be able to be created: the $n - \p$
    replicas that reached $C_j.\mathit{idx}$ with this log leave at most $\p$ correct replicas holding a
    history consistent with $C_{k+1}$, and so at most $\p + f$ \texttt{SYNC} messages for $C_{k+1}$,
    fewer than the $n - \p$ required.

\end{enumerate}

Either case is a contradiction, so $C_j$ is consistent with $C_{j-1}$ and hence with every earlier
checkpoint of the round.
\end{proof}

\begin{corollary}
\label{cor:diverge-bound}
Let $C$ be a checkpoint of round $i$. At any index, at most $\p + f$ replicas can attest --- by
\texttt{SYNC} or by \texttt{SPEC-REPLY} --- to a digest whose history is inconsistent with $C$.
\end{corollary}

\begin{proof}
This follows from the same counting as above. A correct replica holding a history consistent with
$C$ could only leave it by aligning to a checkpoint inconsistent with $C$, and by
Lemma~\ref{lem:ckpt-consistent} no such checkpoint exists in round $i$. So at most $\p$ correct
replicas ever hold a history inconsistent with $C$, and the remaining attestations must come from
the up to $f$ Byzantine replicas.
\end{proof}

\begin{corollary}
\label{cor:ckpt-commit}
If $m$ is fast-path committed at slot $k$ in round $i$, then no checkpoint of round $i$ is
inconsistent with $m$ at $k$. In particular, any checkpoint $C$ of round $i$ with
$C.\mathit{idx} \geq k$ has $m$ at slot $k$.
\end{corollary}

\begin{proof}
The commit is attested by a fast path quorum of $n - \p$ \texttt{SPEC-REPLY} messages carrying a
single digest at $k$. Were some checkpoint $C$ of round $i$ inconsistent with that digest, then by
Corollary~\ref{cor:diverge-bound} at most $\p + f$ replicas could have attested to it, and since
$\p + f < n - \p$ the quorum could not have formed.
\end{proof}

\subsubsection{Agreement Within a Round}

\begin{lemma}
\label{lem:fp-fp}
If $m$ is fast-path committed at slot $k$ in round $i$, then no request $m' \neq m$ can be fast-path
committed at slot $k$ in round $i$.
\end{lemma}

\begin{proof}
For sake of contradiction, suppose two distinct requests $m$ and $m'$ are both fast-path committed at
slot $k$ in round $i$. By definition there exist quorums $S, S'$ of $(n - \p)$ replicas each that
sent \texttt{SPEC-REPLY} for $\langle m, k \rangle$ and $\langle m', k \rangle$, respectively.
 By (Q1) at least $2f + 1$ replicas in $S \cap S'$ are correct. 
 Let $\mathcal{R}$ denote this set of correct replicas in $S \cap S'$.

If $r \in \mathcal{R}$ has sent \texttt{SPEC-REPLY}s for both $m$ and $m'$ at $k$, it must have sent
one before invoking \alignp{} to some checkpoint $C$ and the other after rolling back its state and
completing \alignp{}.
Let $C$ be the checkpoint to which $r$ aligned between issuing its two \texttt{SPEC-REPLY}s. 
Without loss of generality, let $m'$ be the request contained in the post-\alignp{} log.
By construction, every \texttt{SPEC-REPLY} in $S'$ carries a digest consistent with $C$ (since $m'$
is the post-\alignp{} version, computed from $C.\mathit{digest}$), while every \texttt{SPEC-REPLY}
in $S$ carries a digest inconsistent with $C$ (since $m$ is the pre-\alignp{} version of the
divergent log).

Every \texttt{SPEC-REPLY} in $S$ carries the same digest at $k$, and that digest is inconsistent
with $C$. By Corollary~\ref{cor:diverge-bound}, at most $f + \p$ replicas can attest to it. Since
\[
f + \p < n - \p = 3f + \p + 1,
\]
the quorum $S$ of size $n - \p$ cannot exist, contradicting the assumption.

The argument extends to the case where a replica aligns more than once within round $i$,
and so sends more than two \texttt{SPEC-REPLY} messages for an index $k$. For every
pair of requests, the consistency argument applies, using the checkpoint that was aligned
to in order to form the second \texttt{SPEC-REPLY} of the pair.
\end{proof}

\begin{lemma}
\label{lem:fp-rp}
If $m$ is fast-path committed at slot $k$ in round $i$ and \repair{} completes in round $i$ with
repair history $\mathcal{H}$, then the repaired log derived from $\mathcal{H}$ will have
$m$ at slot $k$.
\end{lemma}

\begin{proof}
Let $S$ be the fast-path quorum of $n - \p$ replicas that sent \texttt{SPEC-REPLY} messages for
 $\langle m, k \rangle$, and let $T$ be the set of $n - f$ replicas whose logs comprise
$\mathcal{H}$. Then by (Q2) at least $f + \p + 1$ replicas in $S \cap T$ are correct. Call
this set of correct replicas $\mathcal{R}$.

Each $r \in \mathcal{R}$ must hold $m$ at $k$ when it sends its \texttt{LOG} message: the only way
it would not is if $r$ rolled its state back through $k$ in \alignp{}, but by 
Corollary~\ref{cor:ckpt-commit} no checkpoint of round $i$ can be inconsistent with a message committed
in the fast path. So each $r$ either reports $m$ at $k$, or has $\mathit{chkpt.idx} \geq k$ and a checkpoint
that already places $m$ at $k$.

The repaired log is computed from the highest valid checkpoint in $\mathcal{H}$
(\S\ref{sec:repair:construct}). If that checkpoint's index is at least $k$, its prefix places $m$ at
$k$. Otherwise every $r \in \mathcal{R}$ reports $m$ at $k$, and since the construction rule
preserves any request appearing at the same index with the same digest in $f + \p + 1$ of the
$n - f$ logs, the repaired log has $m$ at slot $k$.
\end{proof}

\begin{lemma}
\label{lem:rp-rp}
At most one repair history is committed in a round $i$.
\end{lemma}

\begin{proof}
\repair{} is a three-phase PBFT-style protocol over a quorum of $n - f$ replicas, parameterized by a
view number $v$ internal to round $i$. We show uniqueness in two steps.

\emph{Within a view.} Any two prepare certificates from the same view $v$ are each composed of
$n - f$ matching \texttt{REPAIR-PREPARE} messages, so by (Q3) at least $2\p + 1 \geq 1$ of their
senders are correct. A correct replica issues at most one \texttt{REPAIR-PREPARE} per view, so the
two certificates must be for the same history.

\emph{Across views.} The argument follows the standard PBFT view-change argument. If a repair
history $\mathcal{H}$ commits in view $v$, then a prepare certificate for $\mathcal{H}$ exists at
some correct replica. The leader of view $v + 1$ collects $n - f$ \texttt{VIEW-CHANGE} messages;
their senders intersect the commit quorum of view $v$ in at least one correct replica, which
forwards the prepare certificate. The view-$v+1$ leader is required to propose a history consistent
with the highest forwarded certificate, so any history committed in view $v + 1$ equals
$\mathcal{H}$. The claim follows by induction on view number.
\end{proof}

\subsubsection{Agreement Across Rounds}

\begin{lemma}
\label{lem:cross-round}
If $m$ is committed at slot $k$ in round $i$, then for every round $i' > i$, no $m' \neq m$ is
committed at slot $k$ in round $i'$.
\end{lemma}

\begin{proof}
Rounds are demarcated by successful \repair{}: if a round $i' > i$ exists, then round $i$ ended with
a successful \repair{} producing some history $\mathcal{H}_i$. By
Lemmas~\ref{lem:fp-fp}--\ref{lem:rp-rp}, every in-round commit at slot $k$ in round $i$ agrees with
$m$; by Lemma~\ref{lem:fp-rp}, $\mathcal{H}_i$ has $m$ at slot $k$. Since a correct replica applies
$\mathcal{H}_i$ before entering round $i + 1$ and records a new checkpoint at the last index of the
repaired log, every correct replica enters round $i + 1$ with $m$ at slot $k$ and with
$\mathit{chkpt.idx} \geq k$ consistent with $m$ at $k$.

We argue by induction on $i' > i$. Assume that every correct replica enters round $i'$ with $m$ at
slot $k$ and $\mathit{chkpt.idx} \geq k$.

\emph{Fast-path commits in round $i'$.} A fast-path commit $\langle m', k \rangle$ requires $n - \p$
matching \texttt{SPEC-REPLY}s with the indicated digest. By the induction hypothesis, every correct
replica's log at $k$ holds $m$. For a correct replica to issue a \texttt{SPEC-REPLY} with a digest
inconsistent with $m$, it must have aligned past $k$ to some checkpoint $C$. As in
Lemma~\ref{lem:fp-fp}, the $n - \p$ \texttt{SYNC} supporters of $C$ contain at least $n - \p - f =
2f + \p + 1$ correct replicas, whose logs (by induction) have $m$ at $k$. Hence $C$ is consistent with $m$ at
$k$ and the post-alignment digest still encodes $m$ at $k$. By the same counting as
Lemma~\ref{lem:fp-fp}, the at most $f + \p$ replicas with a conflicting digest cannot form a
fast-path quorum, so any fast-path commit in round $i'$ commits $m$, i.e., $m' = m$.

\emph{Repair commits in round $i'$.} By the induction hypothesis every correct replica enters round
$i'$ with $\mathit{chkpt.idx} \geq k$, so slot $k$ lies below the start of every correct replica's
uncommitted log and is carried by its checkpoint instead. Since the repaired log is computed from
the highest valid checkpoint in $\mathcal{H}_{i'}$ (\S\ref{sec:repair:construct}), that checkpoint
places $m$ at $k$, and repair leaves the slot untouched.

This establishes the inductive step, and the lemma follows.
\end{proof}

\begin{proof}[Proof of Theorem~\ref{thm:safety}]
Suppose $m$ is committed at slot $k$ in round $i$ and $m'$ at slot $k$ in round $i'$. Without loss
of generality assume $i \leq i'$. If $i = i'$, Lemmas~\ref{lem:fp-fp}--\ref{lem:rp-rp} cover the
three possible combinations of fast-path and repair commits and force $m = m'$. If $i < i'$,
Lemma~\ref{lem:cross-round} forces $m = m'$. Since requests are identified by $(c, s_c)$ and each
commit binds one such identifier to one slot, committed requests map one-to-one onto slots, which
are totally ordered.
\end{proof}

\subsection{Liveness}

We show that after GST, every correct client's request eventually commits. Recall that the network
is partially synchronous: there exists a global stabilization time $\mathrm{GST}$ after which every
message between correct parties is delivered within $\Delta$ time. We assume all clocks run at the
same rate and that local request processing is instantaneous; both simplifications can be relaxed
without affecting the argument.

A client is \emph{correct} if it follows the protocol specification, which includes retransmitting
its request until it commits. Replicas deduplicate by $(c, s_c)$ against requests they
have already committed, so a retransmission of a committed request costs only a reply and no request
is ever appended twice.

We make no assumption about clock synchronization. ETAs determine the order replicas speculatively
execute requests, so poorly synchronized clocks cause divergence and cost the fast path, but they
cannot delay a request indefinitely: a replica gives a request a release deadline at most
$\mathit{maxWait}$ after receipt, measured on its own clock (\S\ref{sec:protocol:etas}). Nor is
any request lost on account of its ETA: \alignp{} discards requests below $\eta^*$ but retains
them as orphans, which it offers to the next \repair{} (\S\ref{sec:protocol:align}). A request
leaves a correct replica only when it commits, or when a \repair{} declines to place it and it is
then too old to requeue (\S\ref{sec:practical}). We assume
$\mathit{TTL} > \Delta + \mathit{maxWait}$.

\begin{theorem}[Liveness]
\label{thm:liveness}
After GST, every request submitted by a correct client eventually commits.
\end{theorem}

We prove Theorem~\ref{thm:liveness} via two lemmas: a completed \repair{} places any request every
correct replica has received (Lemma~\ref{lem:arrival}), and \repair{} is live
(Lemma~\ref{lem:repair-live}).

\begin{lemma}
\label{lem:arrival}
After GST, if a correct client transmits $m$ at time $t \geq \mathrm{GST}$, then every \repair{}
beginning after $t + \Delta + \mathit{maxWait}$ places $m$ in its repaired log.
\end{lemma}

\begin{proof}
Every correct replica receives $m$ by $t + \Delta$ and releases it from $\mathit{etaQ}$ within
$\mathit{maxWait}$ of doing so, so all of them have released $m$ by $t + \Delta + \mathit{maxWait}$.
A released request sits in the replica's log, or in its orphan set if \alignp{} discarded it, and
\repair{} carries both, in $\mathcal{L}$ and $\mathcal{O}$ respectively. Every correct replica
therefore offers $m$, so a \repair{} beginning after this point finds it in at least
$n - 2f = f + 2\p + 1 \geq f + 1$ of the $n - f$ logs of $\mathcal{H}$, which the construction rule
places (\S\ref{sec:repair:construct}). Each correct replica then sends \texttt{COMMITTED-REPLY}.

No correct replica drops $m$ before then: it discards $m$ only at a \repair{} that declines to place
it, and by the above such a \repair{} places $m$ instead.
\end{proof}

\begin{lemma}
\label{lem:repair-live}
After GST, every invocation of \repair{} eventually completes.
\end{lemma}

\begin{proof}
\repair{} is a PBFT-style protocol with leader rotation; we follow the standard liveness argument,
adapted to the round structure of \prot{}.

\emph{Coordinated entry.} A correct replica enters \repair{} only on a \texttt{TIMEOUT-PROOF} of
$f + 1$ \texttt{TIMEOUT} messages or on a \texttt{CONFLICT-PROOF} (\S\ref{sec:repair:trigger});
gathering $n - f$ \texttt{SYNC} messages for an index starts $\mathit{chkptTimeout}$ but does not by itself
trigger entry. Both proofs are self-verifying and are rebroadcast on receipt, so if any correct
replica enters \repair{} at time $t \geq \mathrm{GST}$, every other correct replica enters by
$t + \Delta$.

\emph{Progress under a correct leader.} Suppose the leader of view $v$ is correct and enters its
view at time $t$. The leader gathers $n - f$ \texttt{LOG} messages by $t + 2\Delta$,
broadcasts \texttt{REPAIR-HISTORY} which arrives at correct replicas by $t + 3\Delta$, prepare
exchange completes by $t + 4\Delta$, and commit exchange completes by $t + 5\Delta$. Let $T_v$ be
the view-change timeout for view $v$: the time a replica waits after entering view $v$ before
starting a view change. If $T_v > 5\Delta$, no correct replica triggers a view change before commit.

\emph{Leader rotation.} If the leader of view $v$ is Byzantine and fails to make progress within
$T_v$, every correct replica times out and broadcasts a \texttt{VIEW-CHANGE} message for view $v +
1$. By the standard PBFT view-change procedure, the leader of view $v + 1$ is selected by $r^* = (v
+ 1) \bmod n$. Since at most $f$ leaders can be Byzantine and the rotation visits each replica in
turn, after at most $f + 1$ consecutive view changes a correct leader is reached and, by the
previous paragraph, \repair{} commits within $5\Delta$ of that view's entry. Since $\Delta$ is not
known in advance, replicas double $T_v$ on each failed view, so it eventually exceeds $5\Delta$ and
the argument applies. Correct replicas
participate in view changes regardless of whether they have entered or exited \repair{}, so leader
rotation is not blocked by laggards.
\end{proof}

\begin{proof}[Proof of Theorem~\ref{thm:liveness}]
A correct client multicasts $m$ to all replicas, so after GST every correct replica receives it.
There are three cases.

\emph{Fast path.} If at least $n - \p$ replicas send matching \texttt{SPEC-REPLY} messages for $m$,
the client commits directly.

\emph{Checkpoint.} Byzantine replicas may withhold \texttt{SPEC-REPLY} while still sending
\texttt{SYNC}, so that a checkpoint forms at an index at or beyond $m$'s slot while the client holds
only $n - f - \p$ replies. Essentially $m$ is committed at
the replicas, but the client cannot see it, and no \repair{} is triggered precisely because the
checkpoint succeeded.
The client's retransmission resolves this: a correct replica whose $\mathit{chkpt.idx}$ has passed
$m$'s slot answers with \texttt{COMMITTED-REPLY} carrying $k$ and $H(k)$, and the client commits on
$f + 1$ of them.

\emph{Repair.} Otherwise no checkpoint covering $m$ ever forms. Once $m$'s $\mathit{TTL}$ expires
every correct replica holding it as an orphan releases it into its log (\S\ref{sec:practical}), so
every correct replica has $m$ at some index and must broadcast \texttt{SYNC} for an index at or
beyond it, on reaching a multiple of $I$ or on $\mathit{syncTimeout}$. Since no checkpoint results they
gather a \texttt{TIMEOUT-PROOF} or \texttt{CONFLICT-PROOF} and enter \repair{}. It completes by
Lemma~\ref{lem:repair-live} and places $m$ by Lemma~\ref{lem:arrival}, after which each correct
replica sends \texttt{COMMITTED-REPLY} and the client commits on $f + 1$ of them.
\end{proof}

% The supporting appendices, switched off with \fullappendix above.
\ifdefined\fullappendix
% !TeX root = ../../paper.tex
\section{Baseline Configuration Variants}
\label{sec:appendix:variants}

This appendix presents supporting experiments that were left out of the main body for space. They show the tradeoffs between various configurations of our baseline protocols and informed which configurations we used in \S\ref{sec:evaluation}. We first focus on Autobahn-RSM and the changes we made to Autobahn: we compare Autobahn-RSM
against Autobahn's published artifact and against a variant of Autobahn-RSM with shorter timers, and compare the published artifact against \prot{} with signatures disabled. We also show the effect of
batch size on PBFT and Zyzzyva, and repeat the main comparison with
\SI{16}{\byte} payloads. Unless noted, clients send \SI{1}{\kibi\byte}
payloads, with $n=6$ and $f=1$.

\begin{figure}[!t]
  \centering
  \includegraphics[width=\columnwidth]{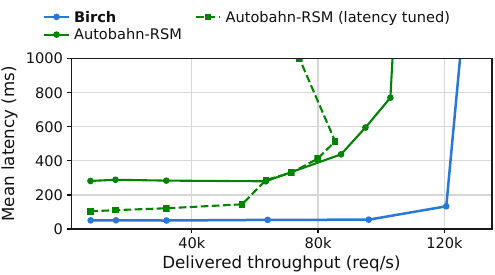}
  \caption{Mean latency against delivered throughput for \prot{} and Autobahn-RSM with its published and latency-tuned timers.}
  \label{fig:variants_autobahn}
\end{figure}

\begin{figure}[!t]
  \centering
  \includegraphics[width=\columnwidth]{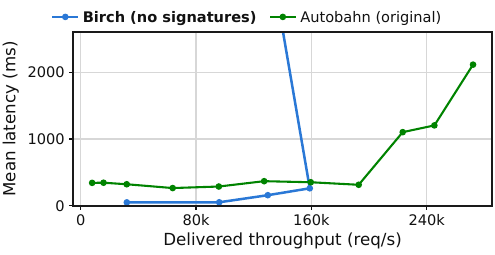}
  \caption{Mean latency against delivered throughput for \prot{} with client signatures disabled and Autobahn's published artifact.}
  \label{fig:variants_nosig}
\end{figure}

\begin{figure}[!t]
  \begin{subfigure}{\columnwidth}
    \centering
    \includegraphics[width=\columnwidth]{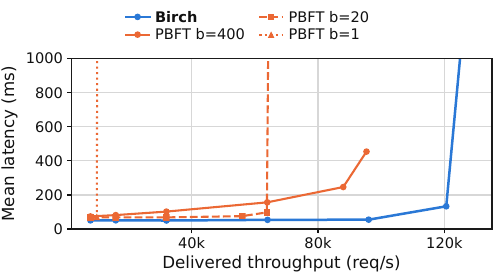}
    \caption{PBFT}
    \label{fig:variants_pbft}
  \end{subfigure}
  \begin{subfigure}{\columnwidth}
    \centering
    \includegraphics[width=\columnwidth]{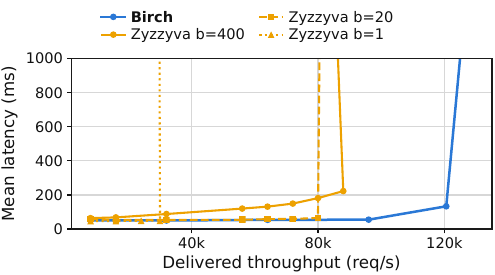}
    \caption{Zyzzyva}
    \label{fig:variants_zyzzyva}
  \end{subfigure}
  \caption{A batch size of $b=20$ shows the best tradeoff between latency and throughput for PBFT and Zyzzyva.}
  \label{fig:variants_batch}
\end{figure}

% Figures for the rest of this appendix and the next, grouped in one float.
\begin{figure*}[!t]
  \begin{minipage}[t]{\columnwidth}
    \vspace{0pt}
    \centering
    \includegraphics[width=\columnwidth]{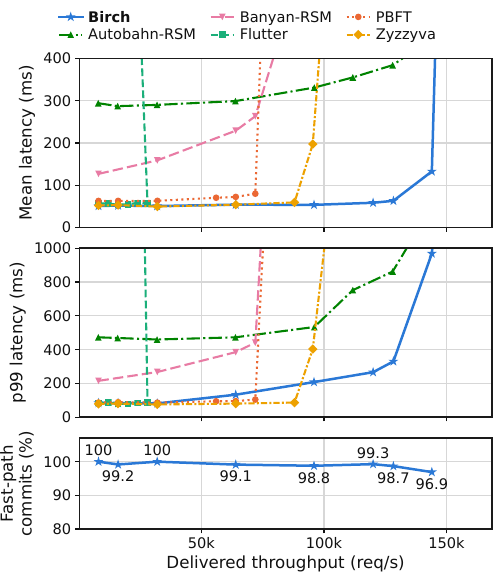}
    \caption{Mean and p99 latency against delivered throughput, and \prot{}'s
      fast path utilization (\SI{16}{\byte}).}
    \label{fig:variants_16b}
  \end{minipage}
  \hfill
  \begin{minipage}[t]{\columnwidth}
    \vspace{0pt}
    \centering
    \includegraphics[width=\columnwidth]{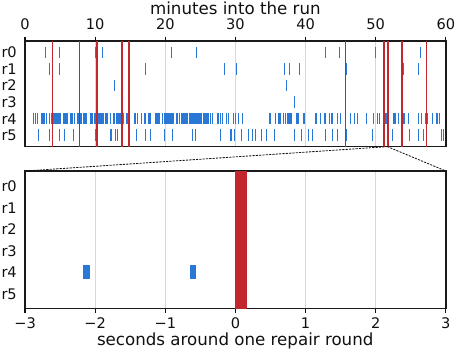}
    \caption{Every \repair{} round (red, across all replicas) and every run of
      \alignp{} (blue, on the replica it realigns) over one hour at 50k
      requests/s (\SI{1}{\kibi\byte}, $n=6$, $f=1$, $\gamma = 1.2$), with one
      \repair{} round blown up below.}
    \label{fig:longrun_timeline}

    \vspace{0.5em}
    \includegraphics[width=\columnwidth]{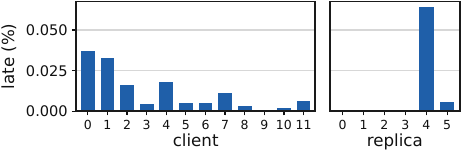}
    \caption{Percentage of requests that arrive at a replica after their ETA,
      by the client that sent them (left) and the replica they arrive at
      (right), over one hour at 50k requests/s.}
    \label{fig:longrun_late_divergence}
  \end{minipage}
\end{figure*}

\subsection{Autobahn Configurations}
\label{sec:appendix:variants:autobahn}
Figure~\ref{fig:variants_autobahn} shows mean latency against delivered throughput for \prot{} and two configurations of Autobahn-RSM. First, we present the same data as used in \S\ref{sec:eval:baselines}, which is Autobahn-RSM with the published configuration. In particular, this configuration sets batch timers to \SI{200}{\milli\second}. However, we also test a \textit{latency-tuned} Autobahn-RSM, which uses shortened batch timers (\SI{5}{\milli\second} each) and enables TCP\_NODELAY on its sockets. This tuning lowers Autobahn-RSM's mean latency under light load from \SI{282}{\milli\second} to \SI{104}{\milli\second}, but also its peak throughput from 103k to 85k requests/s. The two configurations share a maximum batch size of \SI{500}{\kibi\byte} per client; with \SI{200}{\milli\second} timers the batches end up being full at around 30k requests/s aggregate across 12 clients (hence a small decrease in latency initially for the non-tuned line, though this gets largely cancelled out by other latency overheads), but with \SI{5}{\milli\second} timers batches would only be filled at 1.2M requests/s, which Autobahn-RSM never reaches. Thus, the latency-tuned version ends up effectively using smaller batch sizes, explaining its lower peak throughput.

\subsection{Removing Client Authentication}
\label{sec:appendix:variants:nosig}
Figure~\ref{fig:variants_nosig} shows the reverse of the change we made to Autobahn's implementation to create Autobahn-RSM. Instead of adding client signatures to Autobahn, we remove them from \prot{}, and compare against Autobahn's published artifact, which we display as Autobahn (original). This unmodified Autobahn implementation runs on the same machines, but as discussed previously, its clients do not sign requests or receive replies, and its replicas do not verify any per request signatures. Because its clients do not receive replies, its latency measurement is also slightly different, ending when the replica commits a request rather than when the client sees a result. Without the shared client request verification bottleneck, \prot{} reaches a
peak throughput of 159k requests/s, and Autobahn reaches 272k
requests/s. Here, Autobahn's optimizations, including its mempool and client request batching, prove useful in achieving a better peak throughput.

\subsection{PBFT and Zyzzyva Batch Sizes}
\label{sec:appendix:variants:batch}
Figure~\ref{fig:variants_batch} shows PBFT and Zyzzyva with batch sizes of 400,
20 and 1. The main results use a batch size of
20, and these runs are the same as in
Figure~\ref{fig:latency_vs_throughput_baselines}. A batch size of 400 raises peak throughput, to 88k requests/s for both PBFT and Zyzzyva, compared to 64k and 80k with a batch size of 20. This
comes at the cost of higher latency under light load (\SI{74}{\milli\second}
rather than \SI{68}{\milli\second} for PBFT, and \SI{64}{\milli\second} rather
than \SI{58}{\milli\second} for Zyzzyva), and of latency that increases with
load: at 64k requests/s, PBFT's mean latency is \SI{157}{\milli\second} with a
batch size of 400 and \SI{98}{\milli\second} with 20, and at 80k requests/s,
Zyzzyva's is \SI{182}{\milli\second} with 400 and \SI{64}{\milli\second} with
20. Without batching, peak throughput decreases drastically, to 10k requests/s
for PBFT and 30k for Zyzzyva. The unbatched runs predate a change to how our
framework batches signing work, so they are not directly comparable to the
other two batch sizes; within their own sweep, which also measured a batch
size of 50, dropping batching lowers mean latency at 1k requests/s from
\SI{68}{\milli\second} to \SI{61}{\milli\second} for PBFT and from
\SI{53}{\milli\second} to \SI{47}{\milli\second} for Zyzzyva, while peak
throughput falls from 48k requests/s for both. Based on these results, we
chose a batch size of 20 for our main experiments.

\subsection{Smaller Payloads}
\label{sec:appendix:variants:16b}
Figure~\ref{fig:variants_16b} repeats
Figure~\ref{fig:latency_vs_throughput_baselines} with \SI{16}{\byte} payloads.
With smaller payloads, per-request costs such as message processing and
cryptography should dominate over per-byte costs. Overall, this does not
change the picture much: latency under light load remains about the same for
every protocol, and \prot{}, reaching 151k requests/s, still sustains at least $1.4\times$ the throughput of PBFT, Zyzzyva, Flutter, and Banyan-RSM. Peak throughput
increases for every protocol except Flutter, which is still limited by  retry storms. Notably, Banyan-RSM increases its peak throughput substantially, from 24k to 80k requests/s,
and Autobahn-RSM surpasses \prot{}, at 159k requests/s. The fast path utilization is also slightly degraded compared to the \SI{1}{\kibi\byte} experiments.

% !TeX root = ../../paper.tex
\section{Stability Over One Hour}
\label{sec:appendix:longrun}

The experiments in \S\ref{sec:evaluation} last at most 5 minutes, too short
to reliably show how often \repair{} and \alignp{} occur over time, or whether \prot{} can
run for an extended period of time. To resolve these questions, we run \prot{}
for one hour at 50k requests/s with \SI{1}{\kibi\byte} payloads and our 
standard setup: $n=6$, $f=1$, $p=1$ and $\gamma = 1.2$.

\prot{} delivers the offered load for the whole hour, with a mean latency of
\SI{53.0}{\milli\second} and a p99 latency of \SI{89.8}{\milli\second}. $99.84\%$ 
of commits take the fast path, $0.14\%$ are queued behind a \repair{},
and only $0.01\%$ commit in \repair{}. Requests arrive at a replica late relative to their
ETAs $0.012\%$ of the time. Figure~\ref{fig:longrun_late_divergence} shows
that this lateness is not evenly distributed across replicas or across
clients.

Figure~\ref{fig:longrun_timeline} shows when \repair{} and \alignp{} occur during the hour.
In total, \repair{} is only required 10 times. The rounds of \repair{} cluster in time rather than being evenly spaced: 5 rounds run within the first 15 minutes, while none run for 30 minutes in the middle of
the hour. \alignp{} occurs far more often, 335 times, taking a median of
\SI{98}{\milli\second} from a replica detecting that it is behind to being
realigned. It is also concentrated on one replica: 264 of the 335 runs realign
the same replica, which is the one that receives $91\%$ of the late arrivals
(Figure~\ref{fig:longrun_late_divergence}).

Measured at each replica, a \repair{} round takes \SI{145}{\milli\second} on
average. Requests that do not commit on the fast path have a mean latency of
\SI{347}{\milli\second}, longer than the \repair{} itself, since their latency
also includes the time replicas need to detect divergence and reply to the
client.

\FloatBarrier

\fi
% !TeX root = ../../paper.tex
\section{Testbed Measurements}
\label{sec:appendix:testbed_measurements}

\begin{figure}[!t]
  \centering
  \includegraphics[width=\columnwidth]{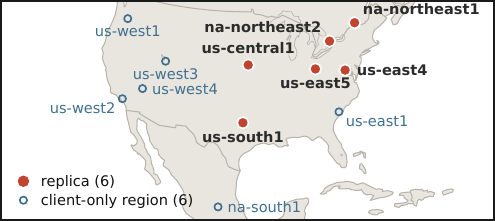}
  \caption{GCP deployment configuration for experiments (unless otherwise noted).}
  \label{fig:deployment_map}
\end{figure}

\begin{figure*}[!tp]
  \centering
  \begin{subfigure}{\textwidth}
    \centering
    \includegraphics[width=\textwidth]{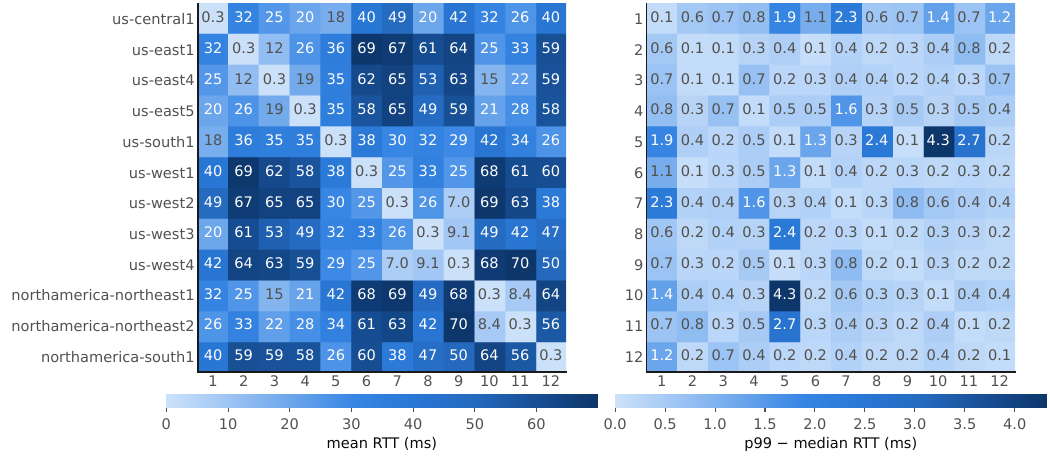}
    \caption{Round trip times (left) and jitter (right).  Each cell is the
      mean of the two directions; columns are numbered as the rows.  Jitter
      is the p99 round trip time above the median, measured within the
      longest single route of each link.}
    \label{fig:matrix-na}
  \end{subfigure}

  \vspace{0.5em}
  \begin{subfigure}{\textwidth}
    \centering
    \includegraphics[width=\textwidth]{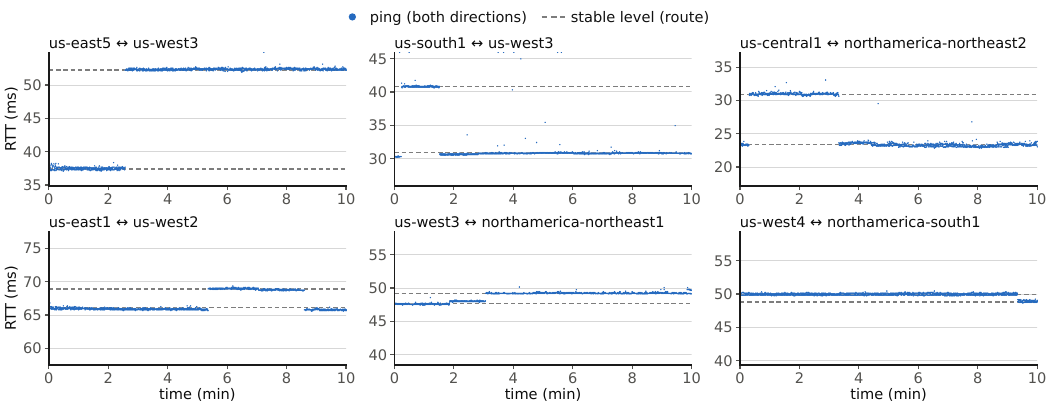}
    \caption{Round trip time over the run for the six links with a detected
      route change: pings in either direction fall into two or more levels,
      each separated by more than \SI{0.5}{\milli\second} and holding at
      least 5\% of the pings.  Dashed lines mark the levels.}
    \label{fig:route-changes}
  \end{subfigure}
  \caption{Measurements between the twelve North American GCP regions, from
    600 ICMP pings per ordered pair over ten minutes.}
  \label{fig:ping-na}
\end{figure*}

Here, we provide more information about our testbed and the measurements we used to create the simulation. Figure~\ref{fig:deployment_map} shows the deployment used for all experiments
unless otherwise noted. Figure~\ref{fig:matrix-na} shows the round trip times and jitter between the
twelve North American GCP regions from which the simulation in
\S\ref{sec:eval:latency} selects replica placements.

These round trip times do not always obey the triangle inequality: for 12 of
the 66 region pairs, relaying through a third region is faster than the direct
link. For example, us-central1 to us-west4 takes \SI{41.7}{\milli\second}
directly but \SI{29.0}{\milli\second} via us-west3, and us-west4 to
northamerica-northeast2 takes \SI{69.9}{\milli\second} directly but
\SI{50.5}{\milli\second} via us-west3.

Figure~\ref{fig:route-changes} shows the six links that changed route during
 the 10 minutes we measured. One of them, us-central1 to northamerica-northeast2,
connects two replicas of our main deployment.

\end{document}